\documentclass[11pt]{article}
\usepackage{latexsym}
\usepackage{amsmath,amsthm,amstext,amscd,amssymb,euscript}
\usepackage{epsfig}
\usepackage{graphicx}
\newtheorem{theorem}{Theorem}[section]
\newtheorem{definition}[theorem]{Definition}
\newtheorem{remark}[theorem]{Remark}

\newtheorem{lemma}[theorem]{Lemma}

\newtheorem{proposition}[theorem]{Proposition}

\newcommand{\beq}{\begin{equation}}
\newcommand{\eeq}{\end{equation}}
\newcommand{\prob}{{\mathbb P}}
\newcommand{\expec}{{\mathbb E}}
\newcommand{\var}{{\mathrm{Var}}}

\newcommand{\ind}{{\large \bf{1}}}

\newcommand{\re}{\ensuremath{\mathcal{R}}}
\newcommand{\aaa}{\ensuremath{\mathcal{A}}}
\newcommand{\spring}{{\medskip}{\noindent}}

\title{A probabilistic approach to Zhang's sandpile model}
\author{Anne Fey-den Boer\footnote{Vrije Universiteit,
De Boelelaan 1081a,
1081 HV Amsterdam, The Netherlands, fey@eurandom.tue.nl, rmeester@few.vu.nl, quant@few.vu.nl} \and Ronald Meester\footnotemark[\value{footnote}]
\and Corrie Quant\footnotemark[\value{footnote}] \and Frank Redig\footnote{Mathematisch Instituut Universiteit Leiden, Niels Bohrweg 1,
2333 CA Leiden,
The Netherlands, redig@math.leidenuniv.nl}}

\begin{document}
\maketitle

\begin{abstract}
The current literature on sandpile models mainly deals with the abelian sandpile model (ASM)
and its variants. We
treat a less known - but equally interesting - model, namely Zhang's sandpile. This
model differs in two aspects from the ASM. First, additions are not discrete, but
random amounts with a uniform distribution on an interval $[a,b]$.
Second, if a site topples - which happens if the amount at that site is larger than a
threshold value $E_c$ (which is a model parameter), then it divides its
entire content in equal amounts among its neighbors. Zhang conjectured that in the
infinite volume limit, this model tends to behave like the ASM in the sense that
the stationary measure for the system in large volumes tends to be peaked
narrowly around a finite set. This belief is supported by simulations, but so far not by analytical
investigations.

We study the stationary distribution of this model in one dimension, for several values of
$a$ and $b$. When there is only one site, exact computations are
possible. Our main result concerns the limit as the number of sites tends to infinity,
in the one-dimensional case. We find that the stationary distribution, in the case $a \geq E_c/2$,
indeed tends to that of the ASM (up to a scaling factor), in agreement with Zhang's
conjecture. For the case $a=0$, $b=1$ we provide strong evidence that the stationary
expectation tends to $\sqrt{1/2}$.
\end{abstract}

\section{Introduction and main results}
With the introduction of the sandpile model by Bak, Tang and Wiesenfeld (BTW), the
notion
of self-organized criticality was introduced, and subsequently applied to several other
models such as forest-fire models, and the Bak-Sneppen model for evolution.
In turn, these models serve as a paradigm for a variety of natural phenomena in which,
empirically, power laws of avalanche characteristics and/or correlations are found,
such as the
Gutenberg-Richter law for earthquakes. See \cite{turcotte} for a extended overview.

After the work of Dhar \cite{dhar}, the BTW model was later renamed `abelian sandpile
model' (ASM), referring to the abelian group structure of addition operators. This
abelianness has since served as the main tool of analysis for this model.

A less known variant of the BTW-model has been introduced by Zhang \cite{zhang}, where
instead
of discrete sand grains, continuous height variables are used. This lattice model is
described informally as follows. Consider a finite subset $\Lambda \subset
\mathbb{Z}^d$.
Initially, every lattice site $i \in \Lambda$ is given an {\em energy} $0\leq E_i
<E_c$, where $E_c$ is the so called {\em critical threshold}, and often chosen to be
equal to 1.
Then, at each discrete time step, one adds a random amount of energy, uniformly
distributed on some interval $[a,b] \subset [0, E_c]$, at a randomly chosen lattice site. If the
resulting energy at this site is
still below the critical value then we have arrived at the new configuration. If not,
an {\em avalanche} is started, in which all unstable sites
(that is, sites with energy at least $E_c$) `topple'
in parallel, i.e., give a fraction $1/2d$ of their energy to each neighbor in
$\Lambda$.
As usual in sandpile models, upon toppling of boundary sites, energy is lost. As in the
BTW-model, the stabilization of an unstable configuration is performed instantaneously,
i.e.,
one only looks at the final stable result of the random addition.

In his original paper, Zhang observes, based on results of numerical simulation (see
also \cite{janosi}), that for large lattices, the energy variables in the stationary
state tend
to concentrate around discrete values of energy; he calls this the emergence of energy
`quasi-units'. Therefore, he argues that in the thermodynamic limit, the stationary
dynamics
should behave as in the discrete ASM. However,
Zhang's model is not abelian (the next configuration depends on the order of topplings
in each avalanche; see below), and thus represents a challenge from the analytical
point of view. There is no mentioning of this fact in \cite{janosi, zhang}, see however \cite{pastor};
probably they chose the usual parallel order of topplings in simulations. 

After its introduction, a model of Zhang's type (the toppling rule is the same as
Zhang's, but the addition is a deterministic amount larger than the critical energy)
has been studied further in the language
of dynamical systems theory in \cite{cessac}. The stationary distributions found for
this model concentrate on fractal sets. Furthermore, in these studies, emergence of
self-organized criticality is linked to the behavior of the smallest
Lyapounov exponents for large system sizes. From the dynamical systems point of view,
Zhang's model is a non-trivial example of an iterated function system, or of a coupled
map lattice with strong coupling.

In this paper we rigorously study Zhang's model in dimension $d=1$ with probabilistic
techniques, investigating uniqueness and deriving certain properties of the stationary
distribution. Without loss of generality, we take $E_c=1$ throughout the paper. In Section
\ref{modeldefsection} we rigorously define the model for $d=1$. We show that in the
particular case of $d=1$ and stabilizing after every addition, the topplings are in
fact abelian, so that the model can be defined without specifying the order of
topplings. In that section, we also include a number of general properties of stationary
distributions. For instance, we prove that if the number of sites is finite, then every
stationary distribution is absolutely continuous with respect to Lebesgue measure on
$(0,1)$, in contrast with the fractal distributions for the model defined in
\cite{cessac} (where the additions are deterministic).

We then study several specific cases of Zhang's model. For each case, we prove by
coupling that the stationary distribution is unique. In Section \ref{onesitesection},
we explicitly compute the stationary distribution for the model on one site, with
$a=0$, by reducing it to the solution of a delay equation \cite{delay}.

Our main result is in Section \ref{halftoteensection}, for the model with $a \geq 1/2$.
We show that in the infinite volume limit, every one-site marginal of the stationary
distribution concentrates on a non-random value, which is the expectation of the
addition distribution (Theorem \ref{quasiunits}). This supports Zhang's conjecture that
in the infinite volume limit, his model tends to behave like the abelian sandpile.
Section \ref{halftoteensection} contains a number of technical results necessary for proving
Theorem \ref{quasiunits}, but which are also of independent interest. For instance, we
construct a coupling of the so-called reduction of Zhang's model to the abelian
sandpile model, and we prove that any initial distribution converges exponentially fast
to the stationary distribution.

In Section \ref{nultoteensection}, we treat the model for $[a,b]=[0,1]$. We present
simulations that indicate the emergence of quasi-units also for this case. However,
since in this case there is less correspondence with the abelian sandpile model, we
cannot fully prove this. We can prove that the stationary distribution is unique, and we
show that if every one-site marginal of the stationary distribution tends to the same
value in the infinite volume limit, and in addition if there is a certain amount of
asymptotic independence, then this value is $\sqrt{1/2}$. This value is consistent with
our own simulations.

\section{Model definition}
\label{modeldefsection}

We define Zhang's model in dimension one as a discrete-time Markov process with state
space $\Omega_N:=
[0, 1)^{ \{1, 2, \ldots, N\} }\subset [0,\infty)^{ \{1, 2, \ldots, N\} }:=\Xi_N$, endowed
with the usual sigma-algebra. We
write $\eta, \xi \in \Omega_N$, configurations of Zhang's model and
$\eta_j$ for the $j$th coordinate of $\eta$. We interpret $\eta_j$ as the amount of
energy at site $j$.

By $\prob_{\eta}$, we denote the probability measure on (the usual sigma-algebra on) the
path space $\Omega_N^\mathbb{N}$ for the process started in $\eta$. Likewise we use
$\prob_{\nu}$ when the process is started from a probability measure $\nu$ on
$\Omega_N$, that is, with initial configuration chosen according to $\nu$.
The configuration at time $t$ is denoted as $\eta(t)$ and its $j$th component as $\eta_j(t)$.

We next describe the evolution of the process. Let $0 \leq a<b\leq 1$. At time 0 the
process starts in some configuration $\eta \in \Omega_N$. For every $t=1,2,\ldots$, the
configuration $\eta(t)$ is obtained from $\eta(t-1)$ as follows.
At time $t$, a random amount of energy $U_{t}$, uniformly distributed on $[a,b]$, is added
to a uniformly chosen site $X_t\in \{ 1,\ldots, N\}$, hence $P(X_{t} = j)=1/N$ for all $j=1,\ldots, N$.
We assume that $U_t$ and
$X_t$ are independent of each other and of the past of the process.
If, after the addition, the energies at all sites are still smaller than 1, then the
resulting
configuration is in $\Omega_N$ and this is the new configuration of the process.

If however after the addition the energy of site $X_t$ is at least 1 - such a site is
called {\em unstable} - then this site will {\em topple}, i.e.,  transfer half of its
energy to its left neighbor and the other half to its right neighbor. In case of a
toppling of a boundary site, this means that half of the energy disappears.
The resulting configuration after one toppling may still not be in $\Omega_N$, because
a toppling may give rise to other unstable sites. Toppling continues until all sites
have energy smaller than 1 (i.e., until all sites are {\em stable}). This final result
of the addition is the new configuration of the process in $\Omega_N$. The entire
sequence of topplings after one addition is called an {\em avalanche}.

We call the above model the $(N, [a,b])$-model. We use the symbol $\mathcal{T}_x(\xi)$
for the result of toppling of site $x$ in configuration $\xi \in \Xi_N$. We write
$\aaa_{u,x}(\eta)$ for the result of adding an amount $u$ at site $x$ of $\eta$,
and stabilizing through topplings.

It is not a priori clear that the process described above is well defined. By this
we mean that it is not a priori clear that every order in which we perform the various
topplings leads to the same final configuration $\eta(t)$. In fact, unlike in the
abelian sandpile, topplings according to Zhang's toppling rule are {\em not} abelian in
general. To give an example of non-abelian behavior, let $N=2$ and $\xi = (1.2, 1.6)$.
Then $\mathcal{T}_1(\mathcal{T}_2(\xi)) = \mathcal{T}_1((2,0)) = (0,1)$, whereas
$\mathcal{T}_2(\mathcal{T}_1(\xi)) = \mathcal{T}_2((0, 2.2)) = (1.1,0)$.

Despite this non-abelianness of certain topplings, we will now show that in the process
defined above, we only encounter avalanches that consist of topplings with the abelian
property. When restricted to a certain subset of $\Omega_N$, topplings are abelian, and
it turns out that this subset is all we use. (In particular, the example that we just
gave cannot occur in our process.)

\begin{proposition}
The $(N, [a,b])$-model is well defined.
\end{proposition}

\begin{proof}
We will prove in two steps that all topplings actually encountered in the process are
abelian. To this end, we first show that in Zhang's model in dimension one, we can
never create two adjacent unstable sites by making one addition to a stable
configuration and toppling unstable sites in any order.

Let $\tilde{\Omega}_N \subset \Xi_N$ be the set of all (possibly unstable)
configurations such that between every pair of unstable sites there is at least one
empty site, and such that the energy of any unstable site is smaller than 2. It is
clear that by making an addition to a stable configuration, we arrive in
$\tilde{\Omega}_N$.
We show that, for every configuration $\tilde{\eta} \in \tilde{\Omega}_N$, the
resulting configuration after toppling of one of the unstable sites is still in
$\tilde{\Omega}_N$.
Introducing some notation, we call a site $j$ of a configuration $\eta$
$$
\begin{array}{lll}
\textrm{empty} & \textrm{if} & \eta_j = 0,\\
\textrm{nonempty} & \textrm{if} & \eta_j \in (0,1),\\
\textrm{unstable} & \textrm{if} & \eta_j \geq 1.\\
\end{array}
$$
An unstable site $i$ of $\tilde{\eta}$ can have either two empty neighbors (first
case), two nonempty neighbors (second case) or one nonempty and one empty (third case).

In the first case, toppling of site $i$ cannot create a new unstable site, since
$\frac 12 \tilde{\eta}_i < 1$, but $i$ itself becomes empty. Thus, if there were
unstable sites to the left and to the right of $i$, after the toppling there still is
an empty site between them.

In the second and third case, the nonempty neighbor(s) of $i$ can become unstable.
Suppose the left neighbor $i-1$ becomes unstable. Directly to its right, at $i$, an empty site is created.
To its left, there was either no unstable site, or first an empty site and then somewhere an
unstable site. The empty site can not have been site $i-1$ itself, because to have
become unstable it must have been nonempty. For the right neighbor the same argument
applies. Therefore, the new configuration is still in $\tilde{\Omega}_N$.

So far, we showed that in the process of stabilization after addition to a stable
configuration, only configurations in $\tilde{\Omega}_N$ are created. In the second
step of the proof we show that, if $\eta \in \tilde{\Omega}_N$ and $i$ and $j$ are
unstable sites in $\eta$, then

\beq
\mathcal{T}_i(\mathcal{T}_j(\eta)) = \mathcal{T}_j(\mathcal{T}_i(\eta)).
\label{topplingsseq}
\eeq
To prove this, we consider all different possibilities for $x$.
If $x$ is not a neighbor of either $i$ or $j$, then toppling of $i$ or $j$ does not
change $\eta_x$, so that (\ref{topplingsseq}) is obvious.
If $x$ is equal to $i$ or $j$, or neighbor to only one of them, then only one of the
topplings changes $\eta_x$, so that again (\ref{topplingsseq}) is obvious.
Finally, if $x$ is a neighbor of both $i$ and $j$, then, since $\eta \in
\tilde{\Omega}_N$, $x$ must be empty before the topplings at $i$ and $j$. We then have
$$
\mathcal{T}_j(\eta)_x = \frac 12 \eta_j,
$$
so that
$$
\mathcal{T}_i(\mathcal{T}_j(\eta))_x = \frac 12 \eta_j + \frac 12 \eta_i =
\mathcal{T}_j(\mathcal{T}_i(\eta))_x
$$
Therefore, also in this last case (\ref{topplingsseq}) is true.

Having established that the topplings of two unstable sites commute, it follows that
the final stable result after an addition is independent of the order in which we
topple, and hence $\aaa_{u,x}(\eta(t))$ is well-defined; see \cite{meester}, Section
2.3 for a proof of this latter fact.
\end{proof}

\begin{remark}
\label{wavedef}
{\em It will be convenient to order the topplings in so-called
{\it waves} \cite{priezzhev}.
Suppose the addition of energy at time $t$ takes place at site $k$ and makes this site
unstable. In the first wave, we topple site $k$ and then all other sites
that become unstable,  {\it but we do not topple site
$k$ again}. After this wave only site $k$ can possibly be unstable. If site $k$ is
unstable after this first wave, the second wave
starts with toppling site $k$ (for the second time) and then
all other sites that become unstable, leaving site $k$ alone, until we reach a
configuration in which all sites
are stable. This is the state of the process at time $t$. It is easy to see that in
each wave, every site can topple at most once.}
\end{remark}

\section{Preliminaries and technicalities}

In this section, we discuss a number of technical results which are needed in the
sequal, and which are also interesting in
their own right. The section is subdivided into three subsections, dealing with
connections to the abelian sandpile, avalanches, and nonsingularity of the marginals of
stationary distributions, respectively.

\subsection{Comparison with the abelian sandpile model}
\label{abeliansection}

We start by giving some background on the abelian sandpile model in one dimension.
In the abelian sandpile model on a finite set $\Lambda \subset \mathbb{Z}$, the amount
of energy added is a nonrandom quantity: each time step one grain of sand is added to a
random site. When a site is unstable, i.e., it contains at least two grains, it topples
by transferring one grain of sand to each of its two neighbors (at the boundary grains
are lost).
The abelian addition operator is as follows: add a particle at site $x$ and stabilize
by toppling unstable sites, in any order.
We denote this operator by $a_x:\{0,1\}^{\Lambda}\to \{0,1\}^{\Lambda}$. For toppling
of site $x$ in the abelian sandpile model, we use the symbol $T_x$.
Abelian sandpiles have some convenient properties \cite{dhar}: topplings on different
sites commute, addition operators commute, and the stationary measure on finitely many
sites is the uniform measure on the set of so-called {\em recurrent} configurations.
Recurrent (or {\em allowed}) configurations are characterized by the fact that they do
not contain a forbidden subconfiguration (FSC). A FSC is defined as a restriction of
$\eta$ to a subset $W$ of $\Lambda$, such that $\eta_x$ is less than the number of
neighbors of $x$ in $W$, for all $x$. In \cite{priezzhev}, a proof can be found that a
FSC cannot be created by an addition or by a toppling.

In the one-dimensional case on $N$ sites, the abelian sandpile model behaves as
follows. Sites are either empty, containing no grains, or full, containing one grain. When
an empty site receives a grain, it becomes full, and when a full site receives a grain,
it becomes unstable. In the latter case, the configuration changes in the following
manner. Suppose the addition site
was $x$. We call the distance to the first site that is empty to the left $i$. If there
is no empty site to the left, then $i-1$ is the distance to the boundary. $j$ is
defined similarly, but now to the right. After stabilization, the sites in
$\{x-i,\ldots,x+j\} \cap \{1,\ldots,N\}$ are full, except for a new empty site at
$x-i+j$. Only sites in  $\{x-i,\ldots,x+j\} \cap \{1,\ldots,N\}$ have toppled.
The amount of topplings of each site is equal to the minimum of its distances to the
endsites of the avalanche. For example, boundary sites can never topple more than once
in an avalanche.
These results follow straightforwardly from working out the avalanche.

The recurrent configurations are those with at most one empty site; a FSC
in the one-dimensional case is a subset of $\Lambda$ of more than 1one site, with
empty sites at its boundary.

Here is an example of how a non-recurrent state on 11 sites relaxes through topplings.
An addition was made to the 7th site; underlined sites are the sites that topple. The
topplings are ordered into waves (see Remark \ref{wavedef}). In the example, the second
wave starts on the 5th configuration:

\begin{eqnarray*}
110111\underline{2}1101
&\to&
11011\underline{2}0\underline{2}101
\to
1101\underline{2}020\underline{2}01
\to
110\underline{2}0121011
\to
111011\underline{2}1011 \\
& \to &
11101\underline{2}0\underline{2}011
\to
1110\underline{2}020111
\to
111101\underline{2}0111
\to
11110\underline{2}01111 \\
&\to&
11111011111.
\end{eqnarray*}

To compare Zhang's model to the abelian sandpile, we label the different states of a
site $j\in\{1,\ldots,N\}$ in $\eta \in \tilde{\Omega}_N$ as follows:

\beq
\begin{array}{lll}
\textrm{empty} (0) & \textrm{if} & \eta_j = 0,\\
\textrm{full} (1) & \textrm{if} & \eta_j \in [\frac 12,1),\\
\textrm{unstable} (2) & \textrm{if} & \eta_j \geq 1,\\
\textrm{anomalous} (a) & \textrm{if} & \eta_j \in (0,\frac 12).\\
\end{array}
\label{reduction}
\eeq

\begin{definition}
The {\em reduction} of a configuration $\eta\in \tilde{\Omega}_N$ is the configuration
denoted by $\re(\eta)\in \{0,1,2,a\}^{\{1,\ldots,N\}}$ corresponding to $\eta$ by
(\ref{reduction}).
\label{reductiondef}
\end{definition}

For general $0 \leq a < b \leq 1$, we have the following result.

\begin{proposition}
For any starting configuration $\eta\in\Omega$, there exists a random variable $T\geq 0$ with $P(T <
\infty)=1$ such that for all $t \geq T$, $\eta(t)$ contains at most one empty or
anomalous site. Moreover, for $t\geq T$, given that $\eta (t)$ contains
an empty site $x_t$, the distribution of this site is uniform
on $\{ 1,\ldots,N\}$.
\label{compare}
\end{proposition}

To prove this proposition, we first introduce FSC's for Zhang's model. We define a FSC
in Zhang's model in one dimension as the restriction of $\eta$ to a subset $W$ of
$\{1,\ldots, N\}$, in such a way that $2\eta_j$ is less than the number of neighbors of
$j$ in $W$, for all $j \in W$. From here on, we will denote the number
of neighbors of $j$ in $W$ by $\mbox{deg}_W(j)$.
To distinguish between the two models, we will from now on
call the above {\em Zhang-FSC}, and the definition given in Section \ref{abeliansection}
{\em abelian-FSC}. From the definition, it follows that a Zhang-FSC in a stable configuration is a restriction to a
subset of more than one site, with the boundary sites either empty or anomalous.
Note that according to this definition, a stable configuration without Zhang-FSC's can be
equivalently described as a configuration with at most one empty or anomalous site.

\begin{lemma}
A Zhang-FSC cannot be created by an addition in Zhang's model.
\label{fsc}
\end{lemma}

\begin{proof}
The proof is similar to the proof of the corresponding fact for abelian-FSC's, which can be
found for instance in \cite{meester}, Section 5.
We suppose that $\eta(t)$ does not contain an FSC, and an addition was made at site
$x$.
If the addition caused no toppling, then it cannot create a Zhang-FSC, because no site
decreased its energy.
Suppose therefore that the addition caused a toppling in $x$. Then for each neighbor
$y$ of $x$
$$
\mathcal{T}_x(\eta)_y \geq \eta_y + \frac 12,
$$
so that $2\mathcal{T}_x(\eta)_y \geq 2\eta_y + 1$. Also $\mathcal{T}_x(\eta)_x = 0$,
and all other sites are unchanged by the toppling.

We will now derive a contradiction. Suppose the toppling created a Zhang-FSC, on a
subset which we call $W$. It is clear that this means that $x$ should be in $W$,
because it is the only site that decreased its energy by the toppling.
For all $j \in W$, we should have that $2\mathcal{T}_x(\eta)_j < \mbox{deg}_W(j)$. This
means that for all neighbors $y$ of $x$ in $W$, we have
$2\eta_y < \mbox{deg}_W(y) - 1$, and for all other $j \in W$ we have
$2\eta_j < \mbox{deg}_W(j)$. From these inequalities it follows that $W\setminus \{x\}$
was already a Zhang-FSC before the toppling, which is not possible, because we supposed
that $\eta(t)$ contained no Zhang-FSC.

By the same argument, further topplings cannot create a Zhang-FSC either, and the proof is complete.
\end{proof}

\begin{remark}
{\em We have not defined Zhang's model in dimension $d>1$, because in that case the
resulting configuration of stabilization through topplings is not independent of the
order of topplings.
But since the proof above only discusses the result of one toppling, Lemma \ref{fsc}
remains valid for any choice of order of topplings. The proof is extended simply by
replacing the factor 2 by $2d$.}
\end{remark}

\medskip\noindent
{\it Proof of Proposition \ref{compare}.}
If $\eta$ already contains at most one non-full, i.e., empty or anomalous site, then it
contains no Zhang-FSC's, and the proposition follows. Suppose therefore that at some
time $t$, $\eta(t)$ contains $M(t)$ non-full sites, with $1<M(t) \leq N$. We denote the
positions of the non-full sites of $\eta(t)$ by $Y_i(t)$, $i = 1, \ldots, M(t)$, and we
will show that $M(t)$ is nonincreasing in $t$, and decreases to 1 in finite time. Note
that for all $1 \leq i<j\leq M(t)$, the restriction of $\eta(t)$ to
$\{Y_i(t),Y_i(t)+1,\ldots,Y_j(t)\}$ is a Zhang-FSC.

At time $t+1$, we have the following two possibilities. Either the addition causes no
avalanche, in that case $M(t+1) \leq M(t)$, or it causes an avalanche. We will call the
set of sites that change in an avalanche (that is, all sites that topple at least once,
together with their neighbors) the {\em range} of the avalanche. We first show that if
the range at time $t+1$ contains a site $y \in \{Y_i(t),\ldots,Y_{i+1}(t)\}$ for some
$i$, then $M(t+1) < M(t)$.

Suppose there is such a site. Then, since $\{Y_i(t)+1,\ldots,Y_{i+1}(t)-1\}$ contains
only full sites, all sites in this subset will topple and after stabilization of this
subset, it will not contain a Zhang-FSC. In other words, in this subset at most one
non-full site is created. But since $Y_i(t)$ and $Y_{i+1}(t)$ received energy from a
toppling neighbor, they are no longer empty or anomalous. Therefore, $M(t+1) < M(t)$.

If there is no such site, then the range is either $\{1,Y_1(t)\}$, or
$\{Y_{M(t)}(t),N\}$. With the same reasoning as above, we can conclude that in these
cases, $Y_1(t+1) < Y_1(t)$, resp. $Y_{M(t)}(t+1) > Y_{M(t)}(t)$.

Thus, $M(t)$ strictly decreases at every time step where an avalanche contains
topplings between two non-full sites. As long as there are at least two non-full sites,
such an avalanche must occur eventually. We cannot make infinitely many additions
without causing topplings, and we cannot infinitely many times cause an avalanche at $x
< Y_1(t)$ or $x > Y_{M(t)}(t)$ without decreasing $M(t)$, since after each such an
avalanche, these non-full sites `move' closer to the boundary.
\qed

\medskip
\noindent In the case that $a \geq 1/2$, we can further specify some
characteristics of the model. We prove that for any initial configuration, after at
most $N(N-1)$ time steps there is at most one empty site, and the other sites are is full, i.e.,
there are no anomalous sites. We will
call such configurations {\em regular}.

\begin{proposition}
Suppose $a \geq \frac12$. Then
\begin{enumerate}
\item
for any initial configuration $\eta$, for all $t \geq N(N-1)$, $\eta(t)$ is regular,
\item
for every stationary distribution $\mu$, and for all $i \in \{1,\ldots,N\}$,
$$
\mu(\eta_i =0) = \frac{1}{N+1}.
$$
\end{enumerate}
\label{compare2}
\end{proposition}
In words, this proposition states that if $a \geq 1/2$, then every
stationary distribution concentrates on regular configurations. Moreover, the
stationary probability that a certain site $i$ is empty, does not depend on $i$. Note
that as a consequence, the stationary probability that all sites are full, is also
$\frac 1{N+1}$.

To prove this proposition, we need the following lemma. In words, it states that if $a
\geq 1/2$ and $\eta$ contains no anomalous sites, then the reduction of Zhang's model
(according to Definition \ref{reductiondef}) behaves just as the abelian sandpile
model.

\begin{lemma}
For all $u \in [\frac12,1)$, for all $\eta\in\Omega_N$ which do not contain anomalous
sites, and for
all $x\in \{ 1,\ldots,N\}$
\begin{equation}
\re (\aaa_{u,x}(\eta))= a_x (\re (\eta)),
\end{equation}
where $a_x$ is the addition operator of the abelian sandpile model.
In both avalanches, corresponding sites topple the same number of times.
\label{comparelemma}
\end{lemma}

\begin{proof}
Under the conditions of the lemma, site $x$ can be either full or empty. If $x$ is
empty, then upon the addition of $u\geq \frac12 $ it becomes full. No topplings follow,
so that in that case we directly have $\re (\aaa_{u,x}(\eta))= a_x (\re (\eta))$.

If $\eta$ is such that site $x$ is full, then upon addition it becomes unstable. We
call the configuration after addition, but before any topplings $\tilde{\eta}$. To
check if in that case $\re (\aaa_{u,x}(\eta))= a_x (\re (\eta))$, we only need to prove
$\re(\mathcal{T}_x(\tilde{\eta})) = T_x(\re (\tilde{\eta}))$, with $\re(\tilde{\eta})_x
= 2$, since we already know that in both models, the final configuration after one
addition is independent of the order of topplings.

In $\mathcal{T}_x(\tilde{\eta})$, site $x$ will be empty. This corresponds to the
abelian toppling, because site $x$ contained two grains after the addition, and by
toppling it gave one to each neighbor.
In $\mathcal{T}_x(\tilde{\eta})$, the energy of the neighbors of $x$ is their energy in
$\eta$, plus at least $\frac 12$.  Thus the neighbors of site $x$ will in
$\mathcal{T}_x(\tilde{\eta})$ be full if they were empty, or unstable if they were
full. Both correspond to the abelian toppling, where the neighbors of $x$ received one
grain.
\end{proof}

\noindent \textit{Proof of Proposition \ref{compare2}}.
To prove part (1), we note that any amount of energy that a site can receive during the
process, i.e., either an addition or half the content of an unstable neighbor, is at
least $1/2$. Thus, anomalous sites can not be created in the process. Anomalous
sites can however disappear, either by receiving an addition, or, as we have seen in
the proof of Proposition \ref{compare}, when they are in the range of an avalanche.

When we make an addition of at least $1/2$ to a configuration with more than one
non-full site, then either the number of non-full sites strictly decreases, or one of
the outer non-full sites moves at least one step closer to the boundary. We note that
$\eta$ contains at most $N$ non-full sites, and the distance to the boundary is at most
$N-1$. When finally there is only one non-full site, then in the next time step it must
either become full or be in the range of an avalanche. Thus, there is a random time
$T'\leq N(N-1)$ such that $\eta(T')$ is regular for the first time, and as anomalous
sites cannot be created, by Proposition \ref{compare}, $\eta(t)$ is regular for all $t
\geq T'$.

For $t \geq T'$, $\eta(t)$ satisfies the condition of Lemma \ref{comparelemma}. This
means that the stationary distribution of the reduction of Zhang's model must coincide
with that of the abelian sandpile model. As we mentioned in Section
\ref{abeliansection}, this is the uniform measure on all configurations with at most
one empty site. This proves part (2).
\qed

\subsection{Avalanches in Zhang's model}

We next describe in full detail the effect of an avalanche, started by an addition to a
configuration $\eta(t)$ in Zhang's model. Let $\mathcal{C}(t+1)$ be the range of this
avalanche. Recall that we defined the range of an avalanche
as the set of sites that change their energy at least once in the course of the
avalanche (that is, all sites that topple at least once, together with their
neighbors). We denote by $\mathcal{T}(t+1)$ the collection of sites that {\em topple} at
least once in the avalanche.
Finally, $\mathcal{C}'(t+1) \subset \mathcal{C}(t+1)$ denotes the collection of anomalous
sites that change, but do not topple in the avalanche.

During the avalanche, the energies of sites in the range, as well as $U_{t+1}$, get
redistributed through topplings in a rather complicated manner. By decomposing the
avalanche into waves (see Remark \ref{wavedef}), we prove the following properties of
this redistribution.

\begin{proposition}\label{belprop}
Suppose an avalanche is started by an addition at site $x$ to configuration $\eta(t)$.
For all sites $j$ in $\mathcal{C}(t+1)$, there exist $F_{ij} =
F_{ij}(\eta(t),x,U_{t+1})$ such that we can write
\beq
\eta_j(t+1) = \sum_{i\in\mathcal{T}(t+1)} F_{ij}\eta_i(t) + F_{xj} U_{t+1} +
\eta_j(t)\ind_{j \in \mathcal{C}'(t+1)},
\label{efjes}
\eeq
with
\begin{enumerate}
\item
\beq
\label{efjes2}
 F_{xj}+ \sum_{i\in\mathcal{T}(t+1)} F_{ij} = \re(\eta(t+1))_j;
\eeq
\item for all $j \in \mathcal{C}(t+1)$ such that $\eta_j(t+1) \neq 0$,
$$
F_{xj} \geq 2^{-\lceil 3N/2\rceil};
$$
\item for all $j \in \mathcal{C}(t+1)$ such that $\eta_j(t+1) \neq 0$, $j \geq x$, we
have
$$
F_{x,j+1} \leq F_{xj};
$$
and similarly, $F_{x,j-1} \leq F_{xj}$ for $j\leq x$.
\end{enumerate}
\label{factortjes}
\end{proposition}

In words, we can write the new energy of each site in the range of the avalanche at
time $t+1$ as a linear combination of energies at time $t$ and the addition $U_{t+1}$,
in such a way that the prefactors sum up to 1 or 0.
Furthermore, every site in the range receives a positive fraction of at least
$2^{-\lceil 3N/2\rceil}$ of the addition. These received fractions are such that larger
fractions are found closer to the addition site. We will need this last property in the
proof of Theorem \ref{quasiunits}.

\medskip\noindent
{\it Proof of Proposition \ref{factortjes}.} We start with part (1). First, we
decompose the avalanche started at site $x$ into waves. We index the waves with $k = 1,
\ldots, K$, and write out explicitly the configuration after wave $k$, in terms of the
configuration after wave $k-1$. The energy of site $i$ after wave $k$ is denoted by
$\tilde{\eta}_{i,k}$; we use the tilde to emphasize that these energies are not really
encountered in the process.
We define $\tilde{\eta}_{i,0} = \eta_i(t)+U_{t+1} \ind_{i=x}$; note that
$\tilde{\eta}_{i,K} = \eta_i(t+1)$.

In each wave, all participating sites topple only once. We call the outermost sites
that toppled in wave $k$, the {\em endsites} of this wave, and we denote them by $M_k$
and $M'_k$, with $M_k > M'_k$. For the first wave, this is either a boundary site, or
the site next to the empty or anomalous site that stops the wave. Thus, $M_1$ and
$M'_1$ depend on $\eta(t)$,
$x$ and $U_{t+1}$. All further waves are stopped by the empty sites that were created
when the endsites of the previous wave toppled, so that for each $k$, $M_{k+1} = M_k
-1$ and $M'_{k+1} = M'_k +1$. In every wave but the last, site $x$ becomes again
unstable. Only in the last wave, $x$ is an endsite, so that at most one of its
neighbors topples.

In wave $k$, first site $x$ topples, transferring half its energy, that is, $\frac 12
\tilde{\eta}_{x,k-1}$, to each neighbor. Then, if $x$ is not an endsite, both its
neighbors topple, transferring half of their current energy, that is, $\frac 12
\tilde{\eta}_{x\pm 1,k-1}+ \frac 14 \tilde{\eta}_{x,k-1}$, to their respective
neighbors. Site $x$ is then again unstable, but it does not topple again in this wave.
Thus, the topplings propagate away from $x$ in both directions, until the endsites are
reached. Every toppling site in its turn transfers half its current energy, including
the energy received from its toppling neighbor, to both its neighbors. Writing out all
topplings leads to the following expression, for all sites $i \geq x$. A similar
expression gives the updated energies for the sites with $i < x$. Note that for every
$k>1$, $\tilde{\eta}_{M_k+1,k-1} = 0$. Only when $k=1$, it can be the case that site
$M_1+1$ was anomalous, so that $\tilde{\eta}_{M_1+1,0} >0$.

\begin{eqnarray}
\label{avalanche}
\tilde{\eta}_{x,k} & = & \left(\frac 12 \tilde{\eta}_{x+1,k-1}+ \frac 14
\tilde{\eta}_{x,k-1}\right)\ind_{M_k>x}+\left(\frac 12 \tilde{\eta}_{x-1,k-1}+ \frac 14
\tilde{\eta}_{x,k-1}\right)\ind_{M'_k<x}, \nonumber\\
\tilde{\eta}_{i,k} & = & \sum_{n=0}^{i+1} \frac{1}{2^{i+2-n}} \tilde{\eta}_{n,k-1},
\mbox{for }
i=x+1,\ldots, M_k-1,\nonumber\\
\tilde{\eta}_{M_k,k} & = & 0,\\
\tilde{\eta}_{M_k+1,k} & = & \left\{ \begin{array}{lll}
\tilde{\eta}_{M_k-1,k}+\tilde{\eta}_{M_k+1,k-1} & \mbox{if}& M_k \geq x+2,\nonumber \\
\frac 12 \tilde{\eta}_{x+1,k-1}+\frac 14 \tilde{\eta}_{x,k-1}+\tilde{\eta}_{x+2,k-1} &
& M_k = x+1,\\
\frac 12 \tilde{\eta}_{x,k-1}+ \tilde{\eta}_{x+1,k-1} & &  M_k = x.\\
\end{array}
\right.
\end{eqnarray}

\medskip\noindent
We write for all $j \in \mathcal{C}(t+1)$, with $f_{ij}(k)$ implicitly defined by the
coefficients in (\ref{avalanche}),
$$
\tilde{\eta}_{j,k} = \sum_{i\in\mathcal{T}(t+1)} f_{ij}(k)\tilde{\eta}_{i,k-1}.
$$
Since we made an addition to a stable configuration, we only encounter configurations
in $\tilde{\Omega}_N$. From a case by case analysis of (\ref{avalanche}), we claim that
for all $j \in \mathcal{C}(t+1)$ we have
\begin{equation}
\label{ronald}
\re(\tilde{\eta}_{j,k}) = \sum_{i\in\mathcal{T}(t+1)}
f_{ij}(k)\re(\tilde{\eta}_{i,k-1});
\end{equation}
the reader can verify this for all cases.

To prove the proposition, we start with $j\in \mathcal{C}'(t+1)$, for which we have
$$
\eta_j(t+1) = \tilde{\eta}_{j,1} = \sum_{i\in\mathcal{T}(t+1)}
f_{ij}(1)\tilde{\eta}_{i,0} + \eta_j(t),
$$
which is \eqref{efjes} with $F_{ij} = f_{ij}(1)$. We also have, according to
(\ref{ronald}),
\begin{eqnarray*}
\re(\eta_j(t+1)) & = & \sum_{i\in\mathcal{T}(t+1)} f_{ij}(1)\re(\tilde{\eta}_{i,0})\\
&=& f_{xj}(1) + \sum_{i\in\mathcal{T}(t+1)} f_{ij}(1),
\end{eqnarray*}
since a site in $\mathcal{C}'(t+1)$ becomes full in the avalanche. This proves part
(\ref{efjes2}) of the proposition for such sites.

For all other sites in $\mathcal{C}(t+1)$, we use induction in $k$.
For wave $k-1$, we make the induction hypothesis that
\beq
\tilde{\eta}_{j,k-1} = \sum_{m\in\mathcal{T}(t+1)} F_{mj}(k-1)\eta_m(t) +
F_{xj}(k-1)U_{t+1},
\label{hyp1}
\eeq
with
\beq
\sum_{m\in\mathcal{T}(t+1)}F_{mj}(k-1)+F_{xj}(k-1) = \re(\tilde{\eta}_{j,k-1}).
\label{hyp2}
\eeq
For $k$ we then obtain
\begin{eqnarray*}
\tilde{\eta}_{j,k} &=& \sum_{i\in\mathcal{T}(t+1)} f_{ij}(k)\tilde{\eta}_{i,k-1}\\
&=& \sum_{m\in\mathcal{T}(t+1)} \sum_{i\in\mathcal{T}(t+1)}F_{mi}(k-1)
f_{ij}(k)\eta_m(t) + \sum_{i\in\mathcal{T}(t+1)}f_{ij}(k) F_{xi}(k-1)U_{t+1}.
\end{eqnarray*}
We also have
\begin{eqnarray*}
\re(\tilde{\eta}_{j,k}) &=& \sum_{i\in\mathcal{T}(t+1)}f_{ij}(k)
\re(\tilde{\eta}_{j,k-1})\\
&=& \sum_{i\in\mathcal{T}(t+1)}f_{ij}(k)\left[
\sum_{m\in\mathcal{T}(t+1)}F_{mi}(k-1)+F_{xi}(k-1)\right].
\end{eqnarray*}
Hence, if we define
$$
F_{mj}(k) = \sum_{i\in\mathcal{T}(t+1)}f_{ij}(k)F_{mi}(k-1),
$$
then (\ref{hyp1}) and (\ref{hyp2}) are also true for wave $k$. For $k-1=0$, the
hypothesis is also true, with $F_{mi}(0) = \ind_{m=i}$.
If we define $F_{ij}:= F_{ij}(K)$, then the first part of the proposition follows.

To prove part (2) of the proposition, we derive a lower bound for $F_{xj}$. The number
$K$ of waves in an avalanche is equal to the minimum of the distance to the end sites,
leading to the upper bound $K \leq \lceil N/2 \rceil$.

After the first wave, (\ref{avalanche}) gives for all nonempty $j \neq x$, $F_{xj}(1)
\geq (\frac 12)^{N+1}$. At the start of the next wave, the fraction of $U_{t+1}$
present at $x$ is equal to $F_{xx}(1) = \frac 12$. Hence, after the second wave, even
if we ignore all fractions of $U_t$ on sites other than $x$, then we still have, again
by (\ref{avalanche}), $F_{xj}(2) > \frac 12(\frac 12)^{N+1}$.
So if before each wave we always ignore all fractions of $U_t$ on sites other than $x$,
and if we assume the maximum number of waves, then we arrive at a lower bound for
nonempty sites $j$:
$$
F_{xj} \geq (\frac 12)^{\lceil N/2\rceil-1} (\frac 12)^{N+1} \geq  2^{-\lceil
3N/2\rceil}.
$$
To prove part (3) of the theorem (we only discuss the case $j\geq x$, since by
symmetry, the case $j \leq x$ is similar), we show that for every $k \in
\{1,\dots,K\}$,
\beq
F_{xx}(k)>F_{x,x+1}(k)> \dots >F_{x,M_k-1}(k) = F_{x,M_k+1}(k),
\label{eis1}
\eeq
and
\beq
F_{x,M_k+1}(k) \geq F_{x,M_{k-1}+1}(k-1).
\label{eis2}
\eeq
This is sufficient to prove the theorem, since for every $k$, site $M_k+1$ does not
change anymore in the waves $k+1, \ldots, K$.

After the first wave, we have from (\ref{avalanche}) that
$$
\frac 12 F_{xx}(1)>F_{x,x+1}(1)> \cdots >F_{x,M_k-1}(1) = F_{x,M_k+1}(1),
$$
so that (\ref{eis1}) and (\ref{eis2}) are satisfied after the first wave.
For all other waves except the last wave, we apply induction.

Assume that after wave $k-1$, for $k<K$,
\beq
\frac 12 F_{xx}(k-1)>F_{x,x+1}(k-1)> \dots >F_{x,M_{k-1}-1}(k-1) =
F_{x,M_{k-1}+1}(k-1).
\label{hyp3}
\eeq
We have seen that this hypothesis is true after the first wave.
We rewrite (\ref{avalanche}), for every $k<K$ (so that $M_k>1$ and
$F_{x,x+1}(k-1) = F_{x,x-1}(k-1)$), as follows:
\begin{eqnarray}
F_{xx}(k) &= & F_{x,x+1}(k-1) + \frac 12 F_{xx}(k-1), \nonumber\\
F_{x,x+1}(k) &= & \frac 12 F_{x,x+2}(k-1) + \frac 14 F_{xx}(k), \nonumber\\
\nonumber F_{x,x+i}(k) &= & \frac 12 F_{x,x+i+1}(k-1) + \frac 12 F_{x,x-i-1}(k)
\hspace{1cm} \mbox{for $i = 2, \ldots, M_k-1$},\\
F_{xM_k}(k) &= & 0,  \nonumber\\
\nonumber F_{x,M_k+1}(k) &= & F_{x,M_k-1}(k).\\
\label{plakjesexpressie}
\end{eqnarray}
From (\ref{plakjesexpressie}) and (\ref{hyp3}), we find the following inequalities,
each one following from the previous one:
$$
F_{xx}(k) = F_{x,x+1}(k-1) + \frac 12 F_{xx}(k-1) < \frac 12 F_{xx}(k-1) + \frac 12
F_{xx}(k-1) = F_{xx}(k-1),
$$
$$
F_{x,x+1}(k) = \frac 12 F_{x,x+2}(k-1) + \frac 14 F_{xx}(k) < \frac 12 F_{x,x+1}(k-1) +
\frac 14 F_{xx}(k-1) = \frac 12 F_{xx}(k),
$$
$$
F_{x,x+2}(k) =  \frac 12 F_{x,x+3}(k-1) + \frac 12 F_{x,x+1}(k) < \frac 12
F_{x,x+2}(k-1) + \frac 14 F_{xx}(k) = F_{x,x+1}(k).
$$
For all $i = 2, \ldots, M_k-1$, if $F_{x,x+i}(k) < F_{x,x+i-1}(k)$ then
\beq
F_{x,x+i+1}(k) =  \frac 12 F_{x,x+i+2}(k-1) + \frac 12 F_{x,x+i}(k) < \frac 12
F_{x,x+i+1}(k-1) + \frac 12 F_{x,x+i-1}(k) = F_{x,x+i}(k).
\label{horinductie}
\eeq
Since $F_{x,x+i}(k) < F_{x,x+i-1}(k)$ is true for $i=2$, (\ref{hyp3}) follows for wave
$k$, and (\ref{eis1}) is proven for every $k<K$.
Moreover, we have
$$
F_{x,M_k+1}(k) =  F_{x,M_k-1}(k) = \frac 12 F_{x,M_k}(k-1) + \frac 12 F_{x,M_k-2}(k).
$$
With the above derived $F_{x,M_k-1}(k) < F_{x,M_k-2}(k)$, it follows that
$F_{x,M_k}(k-1) < F_{x,M_k-2}(k)$,
so that
$$
F_{x,M_k-1}(k) > F_{x,M_k}(k-1) = F_{x,M_{k-1}+1}(k-1),
$$
which is (\ref{eis2}).

Finally we discuss the last wave. In case $M_K = 0$ we have
\begin{eqnarray*}
F_{xx}(K) & =&  0, \\
F_{x,x+1}(K) & = & \frac 12 F_{xx}(K-1) = F_{x,x+2}(K-1). \\
\end{eqnarray*}

In case $M_K = 1$ we have
\begin{eqnarray*}
F_{xx}(K) & = & \frac 12 F_{x,x+1}(K-1) + \frac 14 F_{xx}(K-1), \\
F_{x,x+1}(K) & = & 0,  \\
F_{x,x+2}(K) & = & F_{xx}(K) = \frac 12 F_{x,x+1}(K-1) + \frac 14 F_{xx}(K-1)
>F_{x,x+1}(K-1).\\
\end{eqnarray*}

For all $M_K>1$ we have
\begin{eqnarray*}
F_{xx}(K) & = &\frac 12 F_{x,x+1}(K-1) + \frac 14 F_{xx}(K-1),\\
F_{x,x+1}(K) & = &\frac 12 F_{x,x+2}(K-1)  + \frac 14 F_{xx}(K-1) < F_{xx}(K).\\
\end{eqnarray*}
We can now use (\ref{horinductie}) as above for $i = 1, \ldots, M_K-1$, so that
(\ref{eis1}) and (\ref{eis2}) follow.
\qed

\subsection{Absolute continuity of one-site marginals of stationary distributions}

Consider a one-site marginal $\nu_j$ of any stationary distribution $\nu$ of Zhang's
sandpile model. It is easy to see that $\nu_j$ will have an atom at 0, because after
each avalanche there remains at least one empty site. It is intuitively clear that
there can be no other atoms: by only making uniformly distributed additions, it seems
impossible to create further atoms. Here we prove the stronger statement
that the one-site marginals of any stationary distribution are absolutely continuous
with respect to Lebesgue measure on
$(0,1)$.

\begin{theorem}\label{abscon}
Let $\nu$ be a stationary distribution for Zhang's model on $N$ sites. Every one-site
marginal of $\nu$ is on $(0,1)$ absolutely continuous with respect to Lebesgue measure.
\end{theorem}

\begin{proof}
Let $A \subset (0,1)$ be so that $\lambda(A) = 0$, where $\lambda$ denotes Lebesgue
measure.
We pick a starting configuration $\eta$ according to $\nu$. We define a stopping time
$\tau$ as the
first time $t$ such that all non-zero energies $\eta_i(t)$ contain a nonzero contribution
of at least one
of the added amounts $U_1, U_2,\ldots, U_t$. We then write, for an arbitrary nonzero site
$j$,
\begin{equation}
\label{hallo}
\prob_{\nu}(\eta_j(t)\in A) \leq \prob_{\nu} (\eta_j(t) \in A, \tau <t) + \prob_{\nu}(t
\leq \tau).
\end{equation}
The second term at the right hand side tends to 0 as $t \to \infty$ by
item 2 of proposition \ref{belprop}. We claim that the
first term
at the right hand side is equal to zero. To this end, we first observe that $\eta_j(t)$
is built up of fractions of $\eta_j(0)$ and
the additions $U_1, U_2,\ldots, U_t$. These fractions are random variables themselves,
and
we can bound this term by
\begin{equation}
\label{anne}
\prob_{\nu} \left(\sum_{i=1}^N Z_i \eta_i(0) + \sum_{s=1}^t Y_s U_s \in A, \sum_{s=1}^t
Y_s >0\right),
\end{equation}
where $Z_i$ represents the (random) fraction of $\eta_i(0)$ in $\eta_j(t)$, and $Y_s$
represents
the (random) fraction of $U_s$ in $\eta_j(t)$.

We clearly have that the $U_s$ are all independent of each other and of $\eta_i(0)$ for
all $i$. However, the $U_s$ are not necessarily independent of the $Z_i$ and the $Y_s$,
since the numerical value of the $U_s$ effects the relevant fractions. Also, we know
from the analysis in the previous subsection that the $Z_i$ and $Y_s$ can only take
values in a countable set. Summing over all elements in this set, we rewrite
(\ref{anne}) as
$$
\sum_{z_i, y_s; \sum_s y_s > 0}\prob_{\nu}\left(\sum_{i=1}^N z_i\eta_i(0) +
\sum_{s=1}^t y_s U_s \in A,
Z_i=z_i, Y_s=y_s\right)
$$
which is at most
$$
\sum_{z_i, y_s; \sum_s y_s > 0}\prob_{\nu} \left(\sum_{i=1}^N z_i\eta_i(0) +
\sum_{s=1}^t y_s U_s
\in A\right),
$$
which, by the independence of the $U_s$ and the $\eta_i(0)$, is equal to
$$
\sum_{z_i, y_s; \sum_s y_s > 0} \int \prob_{\nu} \left(\sum_{i=1}^N z_i x_i +
\sum_{s=1}^t y_sU_s\in A\right)
d \nu(x_1,\ldots, x_N).
$$
Since
$\sum_{s=1}^t y_s >0$, $U_s$ are independent uniforms, and by assumption $\lambda (A)=0$,
the probabilities inside the integral are clearly zero.
Since the left hand side of (\ref{hallo}) is equal to $\nu_j(A)$ for all $t$, we now
take the
limit $t \to \infty$ on both sides, and we conclude that $\nu_j(A)=0$.
\end{proof}
\begin{remark}
{\rm The same proof
shows that for every stationary measure
$\nu$, and for every $i_1,\ldots,i_k\in \{ 1,\ldots,N\}$,
conditional on $i_1,\ldots,i_k$ being nonempty, the
joint distribution of $\eta_{i_1},\ldots,\eta_{i_k}$
under $\nu$ is absolutely continuous with respect to
Lebesgue measure on $(0,1)^k$.}
\end{remark}
\section{The $(1,[a,b])$-model}
\label{onesitesection}

In this section we consider the simplest version of Zhang's model: the
$(1,[a,b])$-model. In words: there is only one site and
we add amounts of energy that are uniformly distributed on the interval $[a,b]$, with
$0 \leq a < b\leq 1$.

\subsection{Uniqueness of the stationary distribution}

Before turning to the particular case $a=0$, we prove uniqueness of the stationary
distribution for all $[a,b] \subseteq [0,1]$.

\begin{theorem}
(a) The $(1,[a,b])$ model has a unique stationary distribution $\rho=\rho^{ab}$. For
every initial distribution $\prob_\eta$ on $\Omega_1$, we have time-average total
variation convergence to $\rho$, i.e.,
$$
\lim_{t \to \infty} \sup_{A\subset \Omega_1} \left|\frac 1t \sum_{s=0}^t \prob_{\eta}
\left(\eta(s) \in A\right) - \rho(A)\right| = 0.
$$
(b) In addition, if there exists no integer $m > 1$ such that $[a,b]\subseteq [\frac
1m, \frac1{m-1}]$, (hence in particular if $a=0$), then we have convergence in total
variation to $\rho$ for every initial distribution $\prob_\eta$ on $\Omega_1$, i.e.,
$$
\lim_{t \to \infty} \sup_{A\subset \Omega_1} \left|\prob_{\eta} (\eta(t) \in A) -
\rho(A)\right| = 0.
$$
\label{rho}
\end{theorem}

\begin{proof}
We prove this theorem by constructing a coupling. The two processes to be
coupled have initial configurations $\eta^1$ and $\eta^2$, with $\eta^1$,$\eta^2 \in
\Omega_1$.
We denote by $\eta^1(t)$, $\eta^2(t)$ two independent
copies of the process starting from $\eta^1$ and $\eta^2$ respectively.
The corresponding independent additions at each time step are denoted by $U^1_t$ and
$U^2_t$, respectively.
Let $T_1 = \min\{t: \eta^1(t) = 0\}$ and $T_2 = \min\{t: \eta^2(t)=0\}$.
Suppose (without loss of generality) that $T_2 \geq T_1$.
We define a shift-coupling (\cite{thorisson}, Chapter 5) as follows:
$$
\begin{array}{ll}
\hat{\eta}^1(t) & = ~ \eta^1(t) \hspace{2.6cm} \mbox{for all} ~t,\\
\hat{\eta}^2(t) & = \left\{ \begin{array}{ll} \eta^2(t) & \mbox{for} ~ t < T_2,\\
                                    \eta^1(t - (T_2-T_1)) & \mbox{for} ~t \geq T_2.\\
                                    \end{array} \right.\\
\end{array}
$$
Defining $T = \min\{t: \eta^1(t) = \eta^2(t)=0\}$, we also define the exact coupling
$$
\begin{array}{ll}
\hat{\eta}^1(t) & = ~ \eta^1(t) \hspace{0.7cm} \mbox{for all} ~t,\\
\hat{\eta}^3(t) & = \left\{ \begin{array}{ll} \eta^2(t) & \mbox{for} ~ t < T,\\
                                    \eta^1(t) & \mbox{for} ~t \geq T.\\
                                    \end{array} \right.\\
\end{array}
$$
Since the process is Markov, both couplings have the correct distribution.
We write $\tilde{\prob} = \prob_{\eta_1}\times\prob_{\eta_2}$. Since
$\tilde{\prob}(T_2<\infty) = \tilde{\prob}(T_1<\infty) = 1$, the shift-coupling is
always successful, and (a) follows.

To investigate whether $\eta^1(t) = \eta^2(t)=0$ occurs infinitely often
$\tilde{\prob}$-a.s., we define $\mathcal{N} = \{n: (n-1)a <1 \wedge nb>1\}$; this is
the set of possible numbers of time steps between successive events $\eta^1(t)=0$. In
words, an $n \in \mathcal{N}$ is such that, starting from $\eta^1 = 0$, it is possible
that in $n-1$ steps we do not yet reach energy 1, but in $n$ steps we do. To give an
example, if $a \geq 1/2$, then $\mathcal{N} = \{2\}$.

If the gcd of $\mathcal{N}$ is 1 (this is in particular the case if $a=0$), then the
processes $\{t:\eta^1(t) = 0\}$ and $\{t:\eta^2(t) = 0\}$ are independent aperiodic
renewal processes, and it follows that $\eta^1(t) = \eta^2(t) = 0$ happens infinitely
often $\tilde{\prob}$-a.s.

As we have seen, for $a>0$, the gcd of $\mathcal{N}$ need not be 1. In fact, we can see
from the definition of $\mathcal{N}$ that this is the case if (and only if) there is an
integer $m > 1$ such that $[a,b]\subseteq [\frac 1m, \frac1{m-1}]$. Then $\mathcal{N} =
\{m\}$. For such values of $a$ and $b$, the processes $\{t:\eta^1(t) = 0\}$ and
$\{t:\eta^2(t) = 0\}$ are periodic, so that we do not have a successful exact coupling.
\end{proof}

\subsection{The stationary distribution of the $(1,[0,b])$-model}

We write $\rho^b$ for the stationary measure $\rho^{0b}$ of the $(1,[0,b])$-model and
$F^{b}$ for the distribution function of the amount of energy at stationarity, that
is,
$$
F^b(h) = \rho^b(\eta: 0 \leq \eta \leq h).
$$
We prove the following explicit solution for $F^b(h)$.

\begin{theorem}
\label{onesite}
(a) The distribution function of the energy in the $(1, [0,b])$-model at
stationarity is
given by

\begin{equation} \label{Fthm}
F^b(h) =
\left \{ \begin{array}{ll}
 0  &  \mbox{for} ~h<0, \\
 F^{b}(0) > 0 & \mbox{for} ~h=0, \\
 F^b(0)\sum_{\kappa = 0}^{m_h} \frac{(-1)^{\kappa}}{b^{\kappa} \kappa!}(h- \kappa
b)^{\kappa} e^{\frac{h- \kappa b}{b}} & \mbox{for} ~0 < h \leq 1,
 \\
 1 & \mbox{for} ~h>1,
 \end{array}
      \right.
\end{equation}
where $m_h ={\lceil\frac hb \rceil-1}$ and where
\[
F^b(0) = \frac{1}{\sum_{\kappa = 0}^{m_h} \frac{(-1)^{\kappa}}{b^{\kappa} \kappa!}(1- \kappa
b)^{\kappa} e^{\frac{1- \kappa b}{b}}}
\]
follows from the identity
$F^b(1) =1$.

\medskip\noindent
(b)   For $h \in [0,1]$ we have
$$
\lim_{b \rightarrow 0} F^b(h) = h.
$$
\end{theorem}

\medskip\noindent
We remark that although in (a) we have a more or less explicit expression for $F^b(h)$, the
convergence in (b) is not proved analytically, but rather probabilistically.

\medskip\noindent
{\it Proof of Theorem \ref{onesite}, part (a).}
Observe that the process for one site and $a=0$ is defined as
\beq
\eta(t+1) = \left(\eta(t) + U_{t+1}\right)~ \ind_{\eta(t) + U_{t+1} <1}.
\label{tijdstap}
\eeq
We define $F_t^b= \prob(\eta(t) \leq h)$, and derive an expression for $F_{t+1}^b(h)$
in terms of $F_t^b(h)$. In the stationary situation, these two functions should be
equal. We deduce from (\ref{tijdstap}) that for $0 \leq h \leq 1$,
\beq
 F_{t+1}^b(h) =  \prob(\eta(t)+U_{t+1} \leq h) + \prob(\eta(t) + U_{t+1} \geq 1).
\label{C}
\eeq
We compute for $0 \leq h \leq b,$
\begin{eqnarray}
\nonumber
\prob(\eta(t)+ U_{t+1} \leq h) & = &  \prob(\eta(t)\leq  h-U_{t+1})\\
\nonumber                    & = &  \int_0^h \frac 1b \prob(\eta(t)\leq  h-u) \, du\\
  & = &\label{C2} \int_0^h \frac{1}{b} F_t^b(h-u) \, du,
\end{eqnarray}
and likewise for $b \leq h \leq 1$ we find
\beq
\prob(\eta(t) + U_{t+1} \leq h)  =  \int_{0}^{b}
  \frac{1}{b} (F_t^b(h-u)) \, du.
\label{C3}
\eeq
Finally,
\begin{eqnarray}
\nonumber
\prob(\eta(t)+U_{t+1} \geq 1) & = & \int_0^b \frac{1}{b} (F_t^b(1)-F_t^b(1-u)) \, du \\
 & = & \label{C4}
   \int_{0}^{b} \frac{(1-F^b_t(1-u))}{b} \, du = F_{t+1}^b(0).
\end{eqnarray}

Putting  (\ref{C}), (\ref{C2}), (\ref{C3}) and (\ref{C4})
together leads to the conclusion that the stationary distribution
$F^b(h)$ satisfies
\begin{equation} \label{vwdeF}
F^b(h)=
\left\{ \begin{array}{ll}
\int_{0}^{h} \frac{F^b(h-u)}{b} \, du + F^b(0) & \mbox{if $0 \leq h \leq b$,} \\
\int_{0}^{b} \frac{F^b(h-u)}{b} \, du + F^b(0) & \mbox{if $b \leq h \leq 1$}.
\end{array}
\right.
\end{equation}
Furthermore, since $F^b(h)$ is a distribution function, $F^b(h)=0$ for $h
<0$ and $F^b(1) =1$.
We can rewrite equation (\ref{vwdeF}) as a differential delay equation. We take $f^b(h)
=\frac{d F^b(h)}{dh}$ the density corresponding to $F^b$ for
$0 < h <1$; this density exists according to Theorem \ref{abscon}.

We first consider the case $0<h \leq b$, in which case $m_h=0$.
We differentiate (\ref{vwdeF}) twice, to get
$$
\frac{d f^{b}(h)}{dh} = \frac{1}{b} f^b(h),
$$
which leads to the conclusion that
$F^b(h) = F^b(0)e^{\frac{h}{b}}$, consistent with
(\ref{Fthm}) for $0 \leq h \leq b$.

Now we consider the case $b \leq h \leq 1$.
We differentiate (\ref{vwdeF}) on both sides to get
\begin{equation}
\label{vwdeF2}
f^b(h) = \frac 1b(F^b(h)-F^b(h-b)).
\end{equation}

At this point, we can conclude that the solution is unique and could in principle be
found using the method of steps. However, since we already have the candidate solution
given in Theorem \ref{onesite}, we only need to check that it indeed satisfies equation
(\ref{vwdeF}).
We check that for the derivative $f^b$ of $F^b$ as defined in (\ref{Fthm}), for
$b \leq h \leq 1$,
\begin{eqnarray*}
f^b(h) & = & F^b(0)\sum_{\kappa = 0}^{m_h} (-\frac 1b)^{\kappa} \frac{1}{\kappa!}\left(
\kappa(h- \kappa b)^{\kappa-1}
 e^{\frac{h- \kappa b}{b}}+(h- \kappa b)^\kappa \frac {1}{b} e^{\frac{h- \kappa
b}{b}}\right)\\
 & = & - \frac{F^b(0)}{b}\sum_{\kappa = 0}^{m_h} (-\frac 1b)^{\kappa-1}
\frac{1}{(\kappa-1)!}(h- \kappa b)^{\kappa-1} e^{\frac{h- \kappa b}{b}}\\
 & & + \frac{F^b(0)}{b}\sum_{\kappa = 0}^{m_h} (-\frac 1b)^{\kappa} \frac{1}{\kappa!}
(h- \kappa b)^{\kappa} e^{\frac{h- \kappa b}{b}},
\end{eqnarray*}
whereas
$$
\frac {F^b(h)}{b} =\frac {F^b(0)}{b}\sum_{\kappa = 0}^{m_h} (-\frac 1b)^{\kappa}
\frac{1}{\kappa!} (h- \kappa b)^{\kappa} e^{\frac{h- \kappa b}{b}}
$$
and
\begin{eqnarray*}
-\frac{F^b(h-b)}{b} & = & -\frac{1}{b} \sum_{\kappa = 0}^{m_h-1}
(-\frac 1b)^{\kappa} \frac{1}{\kappa!} \kappa(h- (\kappa+1) b)^{\kappa} e^{\frac{h-
(\kappa+1) b}{b}}
\\
&
= & -\frac{1}{b} F^b(0)\sum_{\kappa = 0}^{m_h} (-\frac 1b)^{\kappa-1}
\frac{1}{(\kappa-1)!}(h- \kappa b)^{\kappa-1} e^{\frac{h- \kappa b}{b}},
\end{eqnarray*}
which leads to (\ref{vwdeF2}) as required.
\qed

\begin{figure}[ht]
 \centerline{\includegraphics[width=7cm]{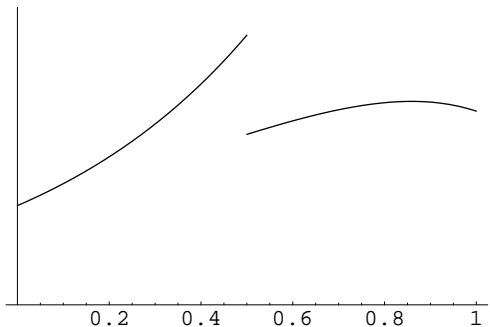}}
 \caption{$f^b(h)$ for $b = \frac 12$. Note the discontinuity at $h = \frac 12$.}
 \label{biseenhalf}
\end{figure}

\begin{figure}[ht]
 \centerline{\includegraphics[width=7cm]{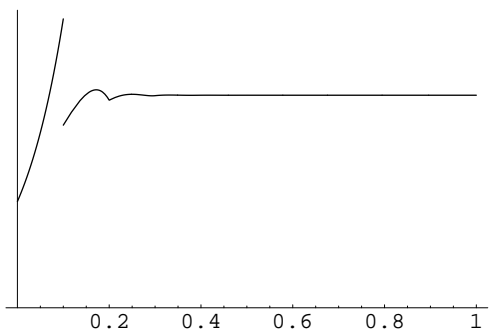}}
 \caption{$f^b(h)$ for $b = \frac 1{10}$. This figure illustrates that for small $b$,
$f^b(h)$ tends to the uniform distribution.}
 \label{biseentiende}
\end{figure}

\medskip\noindent
We remark that the probability
density function $f^b(h)$ has an essential point of discontinuity at $h=b$.
Figures \ref{biseenhalf} and \ref{biseentiende} show two examples of $f^b(h)$.

\medskip
\noindent
{\it Proof of Theorem \ref{onesite}, part (b).}
Without loss of generality, we start at time $0$ with zero energy. To avoid confusion
and express the dependence of the models on the parameter $b$, we will write
$\eta^b(t)$ for the random state of the $(1,[0,b])$-model at time $t$, with $\eta^b=0$.

In this proof, it is helpful to see that the process can be viewed as an alternating
renewal process. To this end, we define the following random times, with $i =
1,2,\ldots$:
$$
T_i^b(0) := \min\{n>T_{i-1}^b(0), \eta^b(n)=0\},
$$
with $T_1^b(0)=0$. We also define
$$
T_i^b(h) := \max\{n \in (T_{i}^b(0),T^b_{i+1}(0)):\eta^b(s) \leq h, \forall s \in
\{T_{i}^b(0)+1,\ldots, n\}\}.
$$
We fix $h$. The process alternates between states where $\eta^b(t)\leq h$ and
$\eta^b(t) >h$, such that renewal events occur at times $T_i^b(0)$. All intervals
between successive $T_i^b(0)$ are i.i.d., as they are all intervals of the form
$T_i^b(h)-T_i^b(0)$. Thus, the requirements of an alternating renewal process are met,
and we can conclude that
\beq
F^b(h) = \lim_{t\to \infty} \prob_{\eta^b}(\eta^b(t)\leq h) = \frac
{\expec(T_1^b(h))+1}{\expec(T_2^b(0))},
\label{altrenewal}
\eeq
which is valid for all $b\leq 1$.

To compute the expectations in (\ref{altrenewal}) in the limit $b \to 0$, we use
another process: let $U_1, U_2, \ldots$ be independent uniform random variables on the
interval $[0,1]$
and write (as before) $S_n^1=U_1+\cdots + U_n$. We define
$$
N(s)= \max \{n \in \mathbb{N}: S_n^1 \leq s \}.
$$
The process $\{N(s):s\geq0\}$ is a renewal process and by the elementary renewal
theorem,
\begin{equation} \label{ert}
\lim_{s \rightarrow \infty}
\frac{\expec(N(s))}{s} = \expec(U_1)=\frac 12.
\end{equation}
Observe that $T_1^b(h)$ and $N(\frac{h}{b})$ have the same distribution; this is
just a rescaling. Likewise, $T_2^b(0)$ has the same distribution as $N(\frac 1b)+1$.
Hence,
\begin{equation} \label{Fb}
F^b(h) = \frac{\mathbb{E}\left( N \left(\frac{h}{b} \right) \right)+1}{\mathbb{E}\left(
N \left(\frac{1}{b} \right) \right)+1},
\end{equation}
and by (\ref{ert}) we find that
$$
\lim_{b \rightarrow 0} \frac{\mathbb{E}\left( N \left(\frac{h}{b} \right) \right)+1}{ \
\frac{h}{2b}} =1.
$$
Combining this with (\ref{Fb}) we conclude that
\begin{eqnarray*}
\lim_{b \rightarrow 0} F^b(h) & = & \lim_{b \rightarrow 0} h \cdot
        \frac{\left(\mathbb{E}\left( N \left(\frac{h}{b} \right) \right) +1 \right)
\cdot \frac{2b}{h}}
             {\left(\mathbb{E} \left( N \left(\frac{1}{b} \right) \right) +1 \right)
\cdot 2b} = h.\\
       \end{eqnarray*}
\qed

\section{The $(N,[a,b])$-model with $N \geq 2$ and $a \geq \frac 12$}
\label{halftoteensection}

\subsection{Uniqueness of stationary distribution}

In the course of the process of Zhang's model, the energies of all sites can be
randomly augmented through additions, and randomly redistributed among other sites
through avalanches. Thus at time $t$, every site contains some linear combination of
all additions up to time $t$, and the energies at time 0. In a series of lemma's we
derive very detailed properties of these combinations in the case $a \geq 1/2$.
These properties are crucial to prove the following result.

\begin{theorem}\label{uniquethm}
The $(N,[a,b])$ model with $a \geq \frac 12$, has a unique stationary
distribution $\mu = \mu^{ab}$.
For every initial distribution $\nu$ on $\Omega_N$, $\prob_\nu$ converges exponentially
fast in total variation to $\mu$.
\label{mu}
\end{theorem}

We have demonstrated for the case $a \geq \frac 12$ that after a finite (random) time,
we only encounter regular configurations (Proposition \ref{compare2}). By Lemma
\ref{comparelemma}, if $\eta(t-1)$ is regular, then the knowledge of $\re(\eta(t-1))$
and $X_t$ suffices to know the number of topplings of each site at time $t$. Thus, also
the factors $F_{ij}$ in Proposition \ref{factortjes} are functions of $\re(\eta(t-1))$
and $X_t$ only. Using this observation, we prove the following.

\begin{lemma}\label{bombom}
Let $a\geq \frac12$. Suppose at some (possibly random) time $\tau$ we have a configuration
$\xi(\tau)$ with no anomalous sites. Then for all $j = 1, \ldots, N$ and for $t \geq
\tau$, we can write
\begin{equation}\label{karamba}
\xi_j(t) = \sum_{\theta=\tau+1}^{t}A_{\theta j}(t)U_{\theta} + \sum_{m=1}^N
B_{mj}(t)\xi_m(\tau)
\end{equation}
in such a way that the coefficients in \eqref{karamba} satisfy
$$
A_{\theta j}(t) = A_{\theta j}(\re(\xi(\tau)),X_{\tau+1},\ldots,X_{t})
$$
and
$$
B_{mj}(t) = B_{mj}(\re(\xi(\tau)),X_{\tau+1},\ldots,X_{t}),
$$
and such that for every $j$ and every $t \geq \tau$,
$$
\sum_{\theta=\tau+1}^{t}A_{\theta j}(t)+\sum_{m=1}^N B_{mj}(t) = \re(\xi(t))_j =
\ind_{\xi_j(t) \neq 0}.
$$
\label{plakjesalgemeen}

\end{lemma}

\begin{remark}
{\em Notice that, in the special case that $\tau$ is a stopping time, $A_{\theta j}(t)$ is independent
of the amounts added after time
$\tau$, i.e., $A_{\theta j}(t)$ and $\{ U_\theta, \theta \geq \tau+1\}$
are independent.
We will make use of this
observation in Section \ref{mainresultssection}.}
\label{plakjesspeciaal}
\end{remark}

\begin{proof}
We use induction. We start at $t=\tau$, where we choose $B_{mj}(\tau)=\re(\xi(\tau))_j~
\ind_{m=j}$. We then have $\sum_{m=1}^N B_{mj}(\tau) = \re(\xi(\tau))_j$, so that at
$t=\tau$ the statement in the lemma is true.
We next show that if the statement in the lemma is true at time $t > \tau$, then it is also true at time
$t+1$.

At time $t$ we have for every $j = 1, \ldots, N$,
$$
\xi_j(t) = \sum_{\theta=\tau+1}^{t}A_{\theta j}(t)U_{\theta} + \sum_{m=1}^N
B_{mj}(t)\xi_m(\tau),
$$
with
$$
\sum_{\theta=\tau+1}^{t}A_{\theta j}(t)+\sum_{m=1}^N B_{mj}(t) = \re(\xi(t))_j,
$$
where all $A_{\theta j}(t)$ and $B_{mj}(t)$ are determined by
$\re(\xi(\tau)),X_{\tau+1},\ldots,X_{t}$, so that $\re(\xi(t))$ is also determined by
$\re(\xi(\tau)),X_{\tau+1},\ldots,X_{t}$. We first discuss the case where we added to a
full site, so that an avalanche is started.
In that case, the knowledge of $\re(\xi(\tau)),X_{\tau+1},\ldots,X_{t+1}$ determines the sets
$\mathcal{C}(t+1)$, $\mathcal{T}(t+1)$ and the factors $F_{ij}$ from Proposition
\ref{factortjes}. (This last fact is due to the fact that $a \geq 1/2$.)
We write, denoting $X_{t+1} = x$,
\begin{eqnarray}
\nonumber \xi_j(t+1) & = & \sum_{i\in\mathcal{T}(t+1)} F_{ij}\xi_i(t) + F_{xj}
U_{t+1}\\
\nonumber & = & \sum_{i\in\mathcal{T}(t+1)} F_{ij}\left[\sum_{\theta=1}^{t}A_{\theta
i}(t)U_{\theta} + \sum_{m=1}^N B_{mi}(t)\xi_m(\tau)\right] + F_{xj} U_{t+1}.\\
\end{eqnarray}
Thus we can identify
\beq
A_{\theta j}(t+1) = \sum_{i\in\mathcal{T}(t+1)} F_{ij} A_{\theta i}(t),
\label{aatjes}
\eeq
$$
B_{mj}(t+1) = \sum_{i\in\mathcal{T}(t+1)} F_{ij} B_{mi}(t),
$$
and
$$
A_{t+1,j}(t+1) = F_{xj},
$$
so that indeed all $A_{\theta j}(t+1)$ and $B_{mj}(t+1)$ are functions of
$\re(\xi(\tau)),X_{\tau+1},\ldots,X_{t+1}$ only. Furthermore,
\begin{eqnarray}
\nonumber
\sum_{\theta=1}^{t+1}A_{\theta j}(t+1) + \sum_{m=1}^N B_{mj}(t+1) & = &
\sum_{i\in\mathcal{T}(t+1)} F_{ij} \left[\sum_{\theta=1}^{t}A_{\theta i}(t) +
\sum_{m=1}^N B_{mi}(t)\right] + F_{xj}\\
\nonumber & = & \sum_{i\in\mathcal{T}(t+1)} F_{ij} + F_{xj} = \re(\eta_j(t+1)),\\
\end{eqnarray}
where we used that all sites that toppled must have been full, therefore had reduced
value 1.

If no avalanche was started, then the only site that changed is the addition site $x$,
and it must have been empty at time $t$. Therefore, we have for all $\theta<t+1$,
$A_{\theta x}(t+1) = A_{\theta x}(t) = 0$, for all $m$, $B_{mx}(t+1) = B_{mx}(t) = 0$
and $A_{t+1,x}(t+1) = 1$, so that the above conclusion is the same.
\end{proof}
\medskip
\noindent For every $\theta$, we have $\sum_{i\in\mathcal{T}(t+1)} A_{\theta i}(t) \leq
1$, because the addition $U_\theta$ gets redistributed by avalanches, but some part
disappears through topplings of boundary sites. One might expect that, as an addition
gets redistributed multiple times and many times some parts disappear at the boundary,
the entire addition eventually disappears, and similarly for the energies
$\xi_j(\tau)$. Indeed, we have the following results about the behaviour of $A_{\theta i}(t)$ for
fixed $\theta$, and about the behaviour of $B_{mj}(t)$ for fixed $m$.

\begin{lemma}
For every $\theta$, and for $t>\theta$,
\begin{enumerate}
  \item $\max_{1\leq i\leq N} A_{\theta i}(t)$ and $\max_{1\leq i\leq N} B_{mi}(t)$ are
both non-increasing in $t$.
  \item For all $\theta, m$ and $i$,
 $\lim_{t \to \infty} A_{\theta i}(t) = 0$, and $\lim_{t \to
\infty} B_{mi}(t) = 0$.
\end{enumerate}
\label{uitsmeren}
\end{lemma}

\begin{proof}
We can assume that $t>\theta$. The proofs for $A_{\theta j}(t)$ and for $B_{mj}(t)$ proceed along the
same line, so we will only discuss $A_{\theta j}(t)$. We will show that for every $j$,
$A_{\theta j}(t+1) \leq \max_i A_{\theta i}(t)$, by considering one fixed $j$. If the
energy of site $j$ did not change in an avalanche at time $t+1$, then
$$
A_{\theta j}(t+1) = A_{\theta j}(t) \leq \max_i A_{\theta i}(t).
$$
If site $j$ became empty in the avalanche, then
$$
A_{\theta j}(t+1) = 0 < \max_i A_{\theta i}(t).
$$
For the third possibility - the energy of site $j$ changed to a nonzero value in an
avalanche at time $t+1$ - we
use \eqref{aatjes}, and estimate
$$
A_{\theta j}(t+1) = \sum_{i\in\mathcal{T}(t+1)} F_{ij} A_{\theta i}(t) \leq \max_i
A_{\theta i}(t) \sum_{i\in\mathcal{T}(t+1)} F_{ij}.
$$
By Proposition \ref{factortjes} part (1) and (2), $\sum_{i\in\mathcal{T}(t+1)} F_{ij}
\leq 1-2^{-\lceil3N/2\rceil}$, so that in this third case,
$$
A_{\theta j}(t+1) \leq (1-2^{-\lceil3N/2\rceil})\max_i A_{\theta i}(t) < \max_i
A_{\theta i}(t).
$$

Thus, it follows that $\max_i A_{\theta i}(t+1)$ can never be larger than $\max_i
A_{\theta i}(t)$.
This proves part (1).
It also follows that when between $t$ and $t+ \tau$ all sites have changed at least
once, we are sure that $\max_i A_{\theta i}(t+\tau) \leq (1-2^{-\lceil3N/2\rceil})\max_i
A_{\theta i}(t)$.

We next derive an upper bound for the time that one of the sites can remain unchanged.
Suppose at some finite time $t$ when $\eta(t)$ is regular (see Proposition
\ref{compare2}), we try never to change some sites again. If all sites are full, then
this is impossible: the next avalanche will change all sites. If there is an empty site
$x$, then the next addition changes either the sites $1,\ldots, x$, or the sites
$x,\ldots, N$.
In the first case, after the avalanche we have a new empty site $x'<x$. If we keep
trying not to change the sites $x,\ldots, N$, we have to keep making additions that
move the empty site closer to the boundary (site 1). It will therefore reach the
boundary in at most $N-1$ time steps. Then we have no choice but to change all sites:
we can either add to the empty site and obtain the full configuration, so that with the
next addition all sites will change, or add to any other site, which immediately
changes all sites.
This argument shows that the largest possible number of time steps between changing all
sites is $N+1$.
We therefore have
\begin{equation}\label{impoa}
\max_i A_{\theta i}(t) <  (1-2^{-\lceil3N/2\rceil})^{\lfloor
\frac{t-\theta}{N+1}\rfloor},
\end{equation}
so that
$$
\lim_{t \to \infty}\max_i A_{\theta i}(t) < \lim_{t \to \infty}
(1-2^{-\lceil3N/2\rceil})^{\lfloor \frac{t-\theta}{N+1}\rfloor} = 0.
$$
\end{proof}
\medskip
\noindent
With the above results, we can now prove uniqueness of the stationary distribution.

\medskip\noindent
{\it Proof of Theorem \ref{mu}.}
By compactness, there is at least one stationary measure $\mu$. To prove the theorem,
we will show that there is a coupling
$(\hat{\eta}^1(t),\hat{\eta}^2(t))_0^{\infty}$ with probability law
$\hat{\prob}_{(\eta^1, \eta^2)}$ for two realizations of the $(N,[a,b])$ model with $a
\geq\frac 12$, such that for all $\epsilon>0$, and for all starting configurations
$\eta^1$ and $\eta^2$, for $t \to \infty$ we have

\beq
\hat{\prob}_{(\eta^1, \eta^2)}\left(\max_j|\hat{\eta}_j^1(t) - \hat{\eta}_j^2(t)| >
\epsilon\right) = O(e^{-\alpha_N t})),
\label{halftoteencoupling}
\eeq
with $\alpha_N > 0$.

From (\ref{halftoteencoupling}), it follows that the Wasserstein distance
(\cite{dudley}, Chapter 11.8) between any two measures $\mu_1$ and $\mu_2$ on
$\Omega_N$ vanishes exponentially fast as $t \to \infty$. If we choose $\eta^1$
distributed according to $\mu$ stationary, then it is clear that every other measure on
$\Omega_N$ converges exponentially fast to $\mu$. In particular, it follows that $\mu$ is unique.

As in the proof of Theorem \ref{rho}, the two processes to be coupled have initial
configurations $\eta^1$ and $\eta^2$, with $\eta^1$,$\eta^2 \in \Omega_N$. The
independent additions at each time step are denoted by $U^1_t$ and $U^2_t$, the
addition sites $X^1_t$ and $X^2_t$.

We define the coupling as follows:
$$
\begin{array}{ll}
\hat{\eta}^1(t) & = \eta^1(t) \hspace{2.8cm} \mbox{for all $t$}\\
\hat{\eta}^2(t) & = \left\{ \begin{array}{ll} \eta^2(t) & \mbox{for} ~ t \leq T,\\
                                    \aaa_{U^1_t,X^1_t}(\hat{\eta}^2(t-1)) & \mbox{for}
~t > T,\\
                                    \end{array} \right.\\
\end{array}
$$
where $T = \min\{t>T':\re(\eta^1(t))=\re(\eta^2(t))\}$, and $T'$ the first time that
both $\eta^1(t)$ and $\eta^2(t)$ are regular. In Proposition \ref{compare2} it was
proven that $T' \leq N(N-1)$, uniformly in $\eta$. In words, this coupling is such that
from the first time on where the reductions of $\hat{\eta}^1(t)$ and $\hat{\eta}^2(t)$
are the same, we make additions to both copies in the same manner, i.e.,
we add the same amounts at the same location to both copies. Then, by Lemma
\ref{comparelemma}, in both copies the same avalanches will occur. We will then use
Lemma \ref{uitsmeren} to show that, from time $T$ on, the difference between
$\hat{\eta}^1(t)$ and $\hat{\eta}^2(t)$ vanishes exponentially fast.

First we show that $\hat{\prob}_{(\eta^1, \eta^2)}(T>t)$ is exponentially decreasing in
$t$. There are $N+1$ possible reduced regular configurations.
Once $\hat{\eta}^1(t)$ is regular, the addition site $X^1_{t+1}$ uniquely determines
the new reduced regular configuration $\re(\eta^1(t+1))$. This new reduced
configuration cannot be the same as $\re(\eta^1(t))$. Thus, there are $N$ equally
likely possibilities for $\re(\eta^1(t+1))$, and likewise for $\re(\eta^2(t+1))$.

If $\re(\eta^1(t)) \neq \re(\eta^2(t))$, then one of the possibilities for
$\re(\eta^1(t+1))$ is the same as $\re(\eta^2(t))$, so that there are $N-1$ possible
reduced configurations that can be reached both from $\eta^1(t)$ and $\eta^2(t)$. The
probability that $\re(\eta^1(t+1))$ is one of these is $\frac{N-1}N$, and the
probability that $\re(\eta^2(t+1))$ is the same is $\frac 1N$. Therefore, $T$ is geometrically distributed, with parameter $p_N = \frac{N-1}{N^2}$.

We now use Lemma \ref{plakjesalgemeen} with $\tau = T$.
For $t>T$, we have in this case that $A^1_{j\theta}(t) = A^2_{j\theta}(t)$ and
$B^1_{jm}(t) = B^2_{jm}(t)$, because from time $T$ on, in both processes the same
avalanches occur. Also, for $t>T$, we have chosen $U^1(t) = U^2(t)$. Therefore, for
$t>T$,
$$
\hat{\eta}_j^1(t) - \hat{\eta}_j^2(t) = \sum_{m=1}^N B^1_{jm}(t)\left(\hat{\eta}^1_m(T)
- \hat{\eta}^2_m(T)\right).
$$
From \eqref{impoa} in the proof of Lemma \ref{uitsmeren} we know that
$$B^1_{jm}(t) \leq (1- 2^{-\lceil 3N/2\rceil})^{\lfloor \frac{t-T}{N+1}\rfloor},$$
so that
$$
\sum_{m=1}^N B^1_{jm}(t)\hat{\eta}^1_m(T) \leq N (1- 2^{-\lceil 3N/2\rceil})^{\lfloor
\frac{t-T}{N+1}\rfloor},
$$
so that for $t>T$, we arrive at
$$
\max_j|\hat{\eta}_j^1(t) - \hat{\eta}_j^2(t)|
\leq 2N (1- 2^{-\lceil 3N/2\rceil})^{\frac{t-T-1}{N+1}}.
$$
We now split $\hat{\prob}_{(\eta^1, \eta^2)}(\max_j|\hat{\eta}_j^1(t) -
\hat{\eta}_j^2(t)|>\epsilon)$ into two terms, by conditioning on $t < 2T$ and $t \geq
2T$ respectively. Both terms decrease exponentially in $t$: the first term because the
probability of $t<2T$ is exponentially decreasing in $t$, and the second term because for
$t \geq 2T$, $\max_j|\hat{\eta}_j^1(t) - \hat{\eta}_j^2(t)|$ itself is exponentially
decreasing in $t$.
\qed

\medskip\noindent
A comparison of the two terms $\prob(t<2T)$ and $\max_j|\hat{\eta}_j^1(t) -
\hat{\eta}_j^2(t)|$ yields that for $N$ large, the second term dominates. We find that
$\alpha_N$ depends, for large $N$, on $N$ as $\alpha_N = -\frac 12 \ln(1- 2^{-\lceil
3N/2\rceil})^{\frac 1{N+1}}$. We see that as $N$ increases, our bound on the speed of
convergence decreases exponentially fast to zero.
Furthermore, we remark that from the above proof it also follows that all stationary
temporal correlations decay exponentially fast, i.e., there exist $\beta_N > 0$ such
that for all square integrable functions $f(\eta)$,
with $\int f^2d\nu=1$, $\int fd\nu =0$,
$$
\expec_\nu\left[ f(\eta(0))f(\eta(t))\right]  \leq
e^{-\beta_N t}.
$$

\subsection{Emergence of quasi-units in the infinite volume limit}
\label{mainresultssection}

In Proposition \ref{compare}, we already noticed a close similarity between the
stationary distribution of Zhang's model with $a \geq 1/2$, and the abelian sandpile
model. We found that the stationary distribution of the reduced Zhang's model, in which
we label full sites as 1 and empty sites as 0, is equal to that of the abelian sandpile
model (Proposition \ref{compare2}).

In this section, we find that in the limit $N \to \infty$, the similarity is even stronger. We find emergence of Zhang's quasi-units in the following sense: as $N \to
\infty$, all one-site marginals of the stationary distribution concentrate on a single,
nonrandom value. We believe that the same is true for $a < 1/2$ also (see Section
\ref{expectedsection} for a related result), but our proof is not applicable in this
case, since it heavily depends on Proposition \ref{compare2}.
To state and prove our result, we introduce the notation $\mu_N$ for the stationary
distribution for the model on $N$ sites, with expectation and variance $\expec^N$ and
$\var^N$, respectively.

\begin{theorem}\label{quasiunits}
In the $(N,[a,b])$ model with $a\geq \frac 12$,
for the unique stationary measure $\mu_N$
we have
\begin{equation}\label{diraccon}
 \lim_{N\to\infty} \mu_N = \delta_{\expec(U)}
\end{equation}
where $\delta_{\expec(U)}$ denotes the Dirac measure
concentrating on the (infinite-volume) constant configuration $\eta_i=\expec (U)$
for all $i\in\mathbb{N}$, and where the limit is in the sense of weak
convergence of probability measures.
\end{theorem}
We will prove this theorem by showing that
for $\eta$ distributed according to
$\mu_N$, in the limit $N \to \infty$, for every sequence $1 \leq j_N \leq N$,
\begin{enumerate}\label{cocoproof}
  \item $\lim_{N \to \infty} \expec^N \eta_{j_N} = \expec U$,
  \item $\lim_{N \to \infty} {\mbox {\em Var}}^N (\eta_{j_N}) = 0$.
\end{enumerate}
\medskip\noindent
The proof of the first item is not difficult. However, the proof of the second part is
complicated, and is split up into several lemma's.

\medskip\noindent
{\it Proof of Theorem \ref{quasiunits}, part (1).}
We choose as initial configuration $\eta \equiv \mathbf{0}$, the configuration with all
$N$ sites empty, so that according to Lemma \ref{plakjesalgemeen}, we can write
\beq
\eta_{j_N}(t) = \sum_{\theta=1}^t A_{\theta j_N}(t)U_{\theta}.
\label{handig}
\eeq
Denoting expectation for this process as $\expec^N_\mathbf{0}$, we find, using Remark
\ref{plakjesspeciaal}, that
$$
\expec^N_\mathbf{0} \eta_{j_N}(t) = \expec U ~\expec^N_\mathbf{0} \re(\eta(t))_{j_N}.
$$
First, we take the limit $t \to \infty$. By Theorem \ref{mu}, $\expec^N_\mathbf{0}\eta_{j_N}(t)$
converges to $\expec^N \eta_{j_N}$. From Proposition \ref{compare2}, it likewise follows that
$\lim_{t \to \infty}\expec^N_\mathbf{0} \re(\eta(t))_j = \frac N{N+1}$.
Inserting these and subsequently taking the limit $ N\to \infty$ proves the first
part.
\qed

\spring
For the proof of the second, more complicated part, we need a number of lemma's. First, we rewrite
$\var^N (\eta_{j_N})$ in the following manner.
\begin{lemma}
$$
{\mbox{\em Var}}^N(\eta_{j_N}) = {\mbox{\em Var}}(U) \lim_{t \to \infty}
\expec^N_\mathbf{0}\left[\sum_{\theta=1}^t(A_{\theta j_N}(t))^2\right] + (\expec U)^2
\frac N{(N+1)^2}.
$$
\label{variantieherschrijven}
\end{lemma}

\begin{proof}
We start from expression (\ref{handig}), and use that the corresponding
variance $\var_\mathbf{0}^N$ converges to the stationary $\var^N$ as $t \to \infty$
by Theorem \ref{uniquethm}. We rewrite, for
fixed $N$ and $j_N = j$,
$$
\var^N_\mathbf{0}(\eta_j(t))  = \expec^N_\mathbf{0}\left[(\eta_j(t))^2\right] -
\left[\expec^N_\mathbf{0} \eta_j(t)\right]^2 =
\expec^N_\mathbf{0}\left[\left(\sum_{\theta=1}^t A_{\theta
j}(t)U_{\theta}\right)^2\right] - \left[\expec^N_\mathbf{0} \sum_{\theta=1}^t A_{\theta
j}(t)U_{\theta}\right]^2
$$
$$
=  \expec^N_\mathbf{0}\left[\sum_{\theta=1}^t (A_{\theta j}(t))^2 U_{\theta}^2 +
\sum_{\theta \neq \theta'} A_{\theta j}(t)U_{\theta}A_{\theta' j}(t)U_{\theta'}\right] -
\left[\expec^N_\mathbf{0} \sum_{\theta=1}^t A_{\theta j}(t)U_{\theta}\right]^2
$$
$$
=  \expec(U^2) \expec^N_\mathbf{0}\left[\sum_{\theta=1}^t (A_{\theta j}(t))^2\right] +
(\expec U)^2\expec^N_\mathbf{0}\left[\sum_{\theta \neq \theta'} A_{\theta
j}(t)A_{\theta' j}(t)\right] - (\expec U)^2\left[\expec^N_\mathbf{0} \sum_{\theta=1}^t
A_{\theta j}(t)\right]^2
$$
$$
 =  \left(\expec(U^2) - (\expec U)^2\right)\expec^N_\mathbf{0}\left[\sum_{\theta=1}^t
(A_{\theta j}(t))^2\right] + (\expec
U)^2\left[\expec^N_\mathbf{0}\left(\sum_{\theta=1}^t A_{\theta j}(t)\right)^2 -
\left(\expec^N_\mathbf{0}\sum_{\theta=1}^t A_{\theta j}(t)\right)^2\right]
$$
$$
=  \var(U)~\expec^N_\mathbf{0}\left[\sum_{\theta=1}^t (A_{\theta j}(t))^2\right] +
(\expec U)^2 \var^N_\mathbf{0} (\re(\eta(t))_j).
$$
Where in the third equality we used the independence of
the $A$-coefficients of the added amounts $U_\theta$.
We now insert $j=j_N$, take the limit $t \to \infty$, and insert $\lim_{t \to \infty}
\var^N_\mathbf{0} (\re(\eta(t))_{j_N}) = \var^N (\re(\eta)_{j_N}) = \frac N{(N+1)^2}$.
\end{proof}

Arrived at this point, in order to prove Theorem \ref{quasiunits}, it suffices to show that
\beq
\lim_{N\to \infty}\lim_{t \to \infty} \expec^N_\mathbf{0}\left[\sum_{\theta=1}^t
(A_{\theta j_N}(t))^2\right]=0.
\label{alleenmaardit}
\eeq
The next lemma's are needed to obtain an estimate for this expectation. We will adopt
the strategy of showing that the factors $A_{\theta j}(t)$ are typically small, so that
the energy of a typical site consists of many tiny fractions of additions. To make this
precise, we start with considering one fixed $\theta > N(N-1)$, a time $t>\theta$, and
we fix $\epsilon>0$.
\begin{definition}
We say that the event $G_t(\alpha)$ occurs, if $\max_j A_{\theta j}(t) \leq
\alpha$. We say that the event $H_t(\epsilon)$ occurs, if  $\max_j A_{\theta j}(t)
\geq \epsilon$, and if in addition there is a lattice interval
of size at most
$M = \lceil \frac 1\epsilon \rceil +1$, containing $X_\theta$, such
that for all sites $j$ outside this interval, $A_{\theta j}(t) \leq \epsilon$.
(We call the mentioned interval the $\theta$-heavy interval.)
\label{zetjes}
\end{definition}

\medskip\noindent
Note that since we have $\sum_j A_{\theta j}(t) \leq 1$ for every $\theta$, the number
of sites where $A_{\theta j}(t) \geq \epsilon$, cannot exceed $\lceil \frac 1\epsilon
\rceil$. In Lemma \ref{uitsmeren}, we proved that $\max_j A_{\theta j}(t)$ is
nonincreasing in $t$, for $t \geq \theta$. Therefore, also $G_t(\alpha)$ is increasing
in $t$. This is not true for $H_t(\epsilon)$, because after an avalanche, the sites
where $A_{\theta j}(t)>\epsilon$ might not form an appropriate interval around $X_{\theta}$.

In view of what we want to prove, the events $G_t(\epsilon)$ and $H_t(\epsilon)$ are
good events, because they imply that (if we think of $N$ as being much larger than $M$)
$A_{\theta j}(t) \leq \epsilon$ `with large probability'.
In the case that $G_t(\epsilon)$ occurs, $A_{\theta i}(t) \leq \epsilon$ for all $i$,
and in the case that $H_t(\epsilon)$ occurs, there can be sites that contain a large
$A_{\theta i}(t)$, but these sites are in the $\theta$-heavy interval containing
$X_\theta$. This latter random variable is uniformly distributed on $\{1,\ldots,N\}$, so that there is a
large probability that a particular $j$ does not happen to be among them. If we only know
that $G_t(\alpha)$ occurs for some $\alpha>\epsilon$, then we cannot draw such a
conclusion. However, we will see that this is rarely the case.

\begin{lemma}
For every $N$, for every $\theta>N(N-1)$, for every $\epsilon>0$, for every $K$ and
$j$,
\begin{enumerate}
\item there exists a constant $c= c(\epsilon)$, such that for $\theta \leq t \leq
\theta+K$
$$
\prob^N_\mathbf{0}(A_{\theta j}(t)>\epsilon) \leq \frac {cK}N;
$$
\item for every $N$ large enough, there exist constants $w=w(\epsilon)$ and
$0<\gamma = \gamma(N,\epsilon)<1$, such that for $t > \theta$
$$
\prob^N_\mathbf{0}(A_{\theta j}(t) > \epsilon) \leq (1-\gamma)^{t-\theta-3w}.
$$
\end{enumerate}
\label{eentheta}
\end{lemma}

\spring
In the proof of Lemma \ref{eentheta}, we need the following technical lemma.
\begin{lemma}
Consider a collection of real numbers $y_i \geq 0$, indexed by $\mathbb{N}$, with $\sum_i y_i
\leq 1$ and such that for some $x\in\mathbb{N}$,
$\max_{i\neq x} y_i \leq \alpha$. Then, for $j \geq x+ \lceil \frac
1{\alpha}\rceil$, we have
$$
\sum_{i=1}^{j-x+2} \frac 1{2^i} y_{j-i+2} ~\leq  ~ f(\alpha) := \left(1- \frac
1{2^{\lceil \frac 1{\alpha} \rceil}}\right)\alpha.
$$
\label{alpha}
\end{lemma}

\begin{proof}
We write
\begin{eqnarray}
\nonumber \sum_{i=1}^{j-x+2} \frac 1{2^i} y_{j-i+2} & =
& \sum_{i=1}^{\lceil \frac 1{\alpha} \rceil} \frac 1{2^i} y_{j-i+2} + \sum_{i= \lceil
\frac 1{\alpha} \rceil +1}^{j-x+2} \frac 1{2^i} y_{j-i+2}\\
\nonumber & \leq & \sum_{i=1}^{\lceil \frac 1{\alpha} \rceil} \frac 1{2^i} y_{j-i+2} +
\frac 1{2^{\lceil\frac 1{\alpha}\rceil +1}} \sum_{i= \lceil \frac 1{\alpha} \rceil
+1}^{j-x+2} y_{j-i+2}.\\
\end{eqnarray}

Note that index $x$ is in the second sum. For $i = 1, \ldots, \lceil\frac
1{\alpha}\rceil$, write
$$
y_{j-i+2} = {\alpha} - z_i,
$$
with $0 \leq z_i \leq {\alpha}$.
Then, since ${\alpha}\lceil\frac 1{\alpha}\rceil \geq 1$ and $\sum_i y_i \leq 1$, we
have $\sum_{i= \lceil\frac 1{\alpha}\rceil +1}^{j-x+2} y_{j-i+2} \leq
\sum_{i=1}^{\lceil\frac 1{\alpha}\rceil} z_i$, so that
$$
\sum_{i=1}^{\lceil \frac 1{\alpha} \rceil} \frac 1{2^i} y_{j-i+2} + \frac
1{2^{\lceil\frac 1{\alpha}\rceil +1}} \sum_{i= \lceil \frac 1{\alpha} \rceil
+1}^{j-x+2} y_{j-i+2}
 \leq \sum_{i=1}^{\lceil \frac 1{\alpha} \rceil} \frac 1{2^i} ({\alpha} - z_i) + \frac
1{2^{\lceil\frac 1{\alpha}\rceil +1}} \sum_{i= 1} ^{\lceil \frac 1{\alpha}\rceil} z_i
$$
$$
= \sum_{i=1}^{\lceil\frac 1{\alpha}\rceil} \frac 1{2^i}{\alpha} -
\sum_{i=1}^{\lceil\frac 1{\alpha}\rceil} \left(\frac 1{2^i} - \frac 1{2^{\lceil\frac
1{\alpha}\rceil +1}}\right) z_i \leq \sum_{i=1}^{\lceil\frac 1{\alpha}\rceil} \frac
1{2^i}{\alpha},
$$
where in the last step we used $z_i \geq 0$. Thus
$$
\sum_{i=1}^{j-x+2} \frac 1{2^i} y_{j-i+2} \leq \sum_{i=1}^{\lceil\frac 1{\alpha}\rceil}
\frac 1{2^i}{\alpha} = \left(1- \frac 1{2^{\lceil \frac 1{\alpha}
\rceil}}\right){\alpha}.
$$
\end{proof}

\spring
{\it Proof of Lemma \ref{eentheta}, part (1).}
We first discuss the case $t=\theta$. We show that $H_\theta(\epsilon)$ occurs, for
arbitrary $\epsilon$.

At $t=\theta$, the addition is made at $X_\theta$. From Proposition \ref{factortjes}
part (3), it follows, for every $\epsilon$, that if after an avalanche there are sites
$j$ with $A_{\theta j}(\theta)>\epsilon$ (we will call such sites `$\theta$-heavy'
sites), then these sites form a set of adjacent sites including $X_\theta$, except for
a possible empty site among them. Since we have $\sum_{j=1}^N A_{\theta j}(\theta) \leq
1$, there can be at most $\lceil \frac 1{\epsilon} \rceil$ $\theta$-heavy sites. If the
addition was made to an empty site, then $A_{\theta X_\theta}(\theta) = 1$. Thus, the
$\theta$-heavy interval has length at most $1+\lceil \frac 1{\epsilon} \rceil$, and we
conclude that $H_\theta(\epsilon)$ occurs.
To estimate the probability that $A_{\theta j}(\theta)>\epsilon$, or in other words,
the probability that a given site $j$ is in the $\theta$-heavy interval, we use that
$X_\theta$ is uniformly distributed on $\{1,\ldots,N\}$. Site $j$ can be in the $\theta-$heavy
interval if the distance between $X_\theta$ and $j$ is at most $1+\lceil
\frac 1{\epsilon} \rceil$, so that $\prob^N_\mathbf{0}(A_{\theta j}(\theta)>\epsilon)
\leq 2\frac{1+\lceil \frac 1{\epsilon} \rceil}N =: \frac {c_2(\epsilon)}N$.

We next discuss $\theta<t\leq \theta+K$. We introduce the following constants. We choose a number $w$ such that $f^w(1) \leq
\epsilon$, with $f$ as in Lemma \ref{alpha}, and where $f^w$ denotes
$f$ composed with itself $w$ times. Note that this is possible because
$\lim_{k \to \infty}f^k(1)=0$. We choose a combination of $\tilde{\epsilon}_0$ and $d$
such that $\tilde{\epsilon}_w \leq \epsilon$, with $\tilde{\epsilon}_{k+1}$ defined as
$\tilde{\epsilon}_k + \frac 1{2^{d+1}}(f^k(1) - \tilde{\epsilon}_k)$.
Finally, we define $\tilde{M_k} = \lceil \frac 1{\tilde{\epsilon}_0} \rceil +1
+k(1+d)$.

For fixed $\theta$ and a time $t > \theta$, we define three
types of avalanches at time $t$: `good', `neutral' and `bad'.

\begin{definition}
For a fixed $\theta$, the avalanche at time $t$ is
\begin{itemize}
\item a {\em good} avalanche if the following conditions are satisfied:
\begin{enumerate}
\item $X_t$ and $X_\theta$ are on the same side of the empty site (if present) at
$t-1$,
\item $X_\theta$ is at distance at least $\tilde{M}_w$ from the boundary,
\item $X_t$ is at distance at least $w$ from the boundary, and from the empty site (if
present) at $t-1$,
\item $X_t$ is at distance at least $\tilde{M}_w+ \lceil \frac 1\epsilon \rceil$ from
$X_\theta$,
\item $X_\theta$ is at distance at least $\tilde{M}_w$ from the empty site (if present)
at $t-1$,
\end{enumerate}
\item a {\em neutral} avalanche if condition (5) is satisfied, but (1) is not,
\item a {\em bad} avalanche in all other cases.
\label{goodavalanche}
\end{itemize}
\end{definition}

Having defined the three kinds of avalanches, we now claim the following:
\begin{itemize}
\item If $H_{t-1}(\epsilon)$ occurs, then after a neutral avalanche, $H_t(\epsilon)$
occurs.
\item If $H_{t-1}(\epsilon)$ occurs, then after a good avalanche, $G_t(\epsilon)$
occurs.
\end{itemize}

The first claim (about the neutral avalanche) holds because if condition (5) is satisfied, but
(1) is not, then not only is $X_\theta$ not in the range of the avalanche, but the
distance of $X_\theta$ to the empty site is large enough to guarantee that the entire
$\theta$-heavy interval is not in the range. Thus, the $\theta$-heavy interval does not
topple in the avalanche. It then automatically follows that $H_t(\epsilon)$ occurs.

To show the second claim (about the good avalanche), more work is required. We break the
avalanche up into waves.
Using a similar notation as in the proof of Proposition \ref{factortjes}, we will
denote by $\tilde{A}_{\theta j}(k)$ the fraction of $U_\theta$ at site $j$ after wave
$k$. We also define another event:
we say that $\tilde{H}_k(\tilde{M}_k,\tilde{\alpha_k},\tilde{\epsilon_k})$ occurs, if
$\max_j \tilde{A}_{\theta j}(k)\leq \tilde{\alpha}(k)$, and all sites where
$\tilde{A}_{\theta j}(k) > \epsilon$ are in an interval of length at most $\tilde{M}_k$
containing $X_\theta$ (we will call this the $(k$-$\theta)$-heavy interval), with the
exception of site $X_t$ when it is unstable, in which case we require that
$\tilde{A}_{\theta X_t}(k) \leq 2\tilde{\epsilon}_k$.

We define a `good' wave, as a wave in which all sites of the $(k$-$\theta)$-heavy interval
topple, and the starting site $X_t$ is at a distance of at least $\frac 1{\alpha_k}$
from the $(k$-$\theta)$-heavy interval. It might become clear now that Definition
\ref{goodavalanche} has been designed precisely so that a good avalanche is an avalanche that
starts with at least $w$ good waves. We will now show by induction in the number of
waves that after an avalanche that starts with $w$ good waves, $G_t(\epsilon)$ occurs.

For $k=0$, we choose $\tilde{\alpha}_0 = 1$, so that at $k=0$,
$\tilde{H}_0(\tilde{M}_0,\tilde{\alpha_0},\tilde{\epsilon_0})$ occurs. We will choose
$\tilde{\alpha}_{k+1} = f(\tilde{\alpha}_k)$ and $\tilde{\epsilon}_{k+1} =
\tilde{\epsilon}_k + \frac 1{2^{d+1}} (\tilde{\alpha}_k-\tilde{\epsilon}_k)$, so that
once $\tilde{H}_w(\tilde{M}_w,\tilde{\alpha_w},\tilde{\epsilon_w})$ occurred, we are
sure that after the avalanche $G_t(\epsilon)$ occurs, because both $\tilde{\alpha_w}$
and $\tilde{\epsilon_w}$ are smaller than $\epsilon$. Now all we need to show is that,
if $\tilde{H}_k(\tilde{M}_k,\tilde{\alpha_k},\tilde{\epsilon_k})$ occurs, then after a
good wave
$\tilde{H}_{k+1}(\tilde{M}_{k+1},\tilde{\alpha}_{k+1},\tilde{\epsilon}_{k+1})$ occurs.

From (\ref{avalanche}), we see that, for all $j>X_t$ that topple (and do not become
empty),
\beq
\tilde{A}_{\theta j}(k+1) = \frac 12 \tilde{A}_{\theta, j+1}(k) + \frac 14
\tilde{A}_{\theta j}(k) + \frac 18 \tilde{A}_{\theta, j-1}(k)+ \cdots + \frac
1{2^{j-X_t+2}} \tilde{A}_{\theta, X_t}(k),
\label{jisnietx}
\eeq
and similarly for $j<X_t$.
For $j=X_t$, we have
\beq
\tilde{A}_{\theta j}(k+1) = (\frac 12 \tilde{A}_{\theta, j+1}(k)+ \frac 14
\tilde{A}_{j,\theta}(k))\ind_{\tilde{A}_{\theta, j+1}(k)\neq 0}+(\frac 12
\tilde{A}_{\theta, j-1}(k)+ \frac 14 \tilde{A}_{j,\theta}(k))\ind_{\tilde{A}_{\theta,
j-1}(k)\neq 0}.
\label{jisx}
\eeq

First we use that in a good wave, all sites in the $\theta$-heavy interval topple, and
$X_t$ is not in this interval. We denote by $m$ the leftmost site of the $\theta$-heavy
interval, so that the rightmost site is $m+\tilde{M}(k)$.
Suppose, without loss of generality, that $X_t<m$. We substitute $\tilde{A}_{\theta
j}(k) \leq \tilde{\alpha}_k$ for all $j$ in the $\theta$-heavy interval,
$\tilde{A}_{X_t \theta}(k) \leq 2\tilde{\epsilon}_k$, and $\tilde{A}_{\theta j}(k) \leq
\tilde{\epsilon}_k$ otherwise into \eqref{jisnietx}, to derive for all $j$ that topple:
\beq
\begin{array}{ll}
j < m-1, j \neq x & \tilde{A}_{\theta j}(k+1) \leq \tilde{\epsilon}_k,\\
j = m-1, \ldots, m+\tilde{M}(k) & \tilde{A}_{\theta j}(k+1) < \tilde{\alpha}_k,\\
j = m+\tilde{M}(k)+d' & \tilde{A}_{\theta j}(k+1) < \tilde{\epsilon}_k + \frac
1{2^{d'+1}} (\tilde{\alpha}_k-\tilde{\epsilon}_k), \hspace{1cm} d'=1,2,\ldots\\
\end{array}
\label{goodwave}
\eeq
Additionally, by (\ref{jisx}), we have $\tilde{A}_{\theta, X_t}(k+1) \leq
2\tilde{\epsilon}_k$. The factor 2 is only there as long as site $X_t$ is unstable.
From \eqref{goodwave} we have that $\tilde{\alpha}_{k+1} < \tilde{\alpha}_k$, but
moreover, in a good wave, the variables $\tilde{A}_{\theta j}(k)$ satisfy the
conditions of Lemma \ref{alpha}, so that in fact $\tilde{\alpha}_{k+1} \leq
f(\tilde{\alpha}_k)$.
If we insert our choice of $d$ for $d'$, then we can see from \eqref{goodwave} that
indeed after the good wave
$\tilde{H}_{k+1}(\tilde{M}_{k+1},\tilde{\alpha}_{k+1},\tilde{\epsilon}_{k+1})$ occurs.

Now we are ready to evaluate $\prob^N_\mathbf{0}(A_{\theta j}(t)>\epsilon)$, for $t \in
\{\theta+1, \ldots, \theta+K\}$. As is clear by now, there are three possibilities:
$G_t(\epsilon)$ occurs, so that $\max_jA_{\theta j}(t) \leq \epsilon$, or
$H_t(\epsilon)$ occurs, in which case $A_{\theta j}(t)$ can be larger than $\epsilon$
if $j$ is in the $\theta$-heavy interval. We derived in the case $t=\theta$ that the
probability for this is bounded above by $\frac {c_2}N$. Finally, it is possible that
neither occurs, in which case we do not have an estimate for the probability that
$A_{\theta j}(t) > \epsilon$. But for this last case, we must have had at least one bad
avalanche between $\theta+1$ and $t$. We will now show that the probability of this
event is bounded above by $\frac{K c_1}N$, where $c_1$ depends only on $\epsilon$.

As stated in Definition \ref{goodavalanche}, a bad avalanche can occur at time $t$ if
at least one of the conditions (2) through (5) is not satisfied. Thus, we can bound
the total probability of a bad avalanche at time $t$, by summing the probabilities that
the various conditions are not satisfied. We discuss the conditions one by one.
\begin{itemize}
\item The probability that condition (2) is not satisfied, is bounded above by
$\frac{2\tilde{M}_w}N$, since $X_\theta$ is distributed uniformly on $\{1,\ldots,N\}$.
\item The probability that condition (3) is not satisfied, is bounded above by
$\frac{4w}N$, since $X_t$ is distributed uniformly on $\{1,\ldots,N\}$, and independent
of the position of the empty site at $t-1$, if present.
\item The probability that condition (4) is not satisfied, is bounded above by
$\frac{2(\tilde{M}_w+\lceil\frac 1\epsilon \rceil)}N$, since $X_t$ and $X_\theta$ are
independent.
\item The probability that condition (5) is not satisfied is bounded above by
$\frac{2\tilde{M}_w}N$, since the position of the empty site at $t-1$ is uniform on
$\{1,\ldots,N\}$.
\end{itemize}
Thus, the total probability of a bad avalanche at time $t$ is bounded by
$\frac{2\tilde{M}_w}N + \frac{4w}N + \frac{2(\tilde{M}_w+\lceil\frac 1\epsilon
\rceil)}N + \frac{2\tilde{M}_w}N \equiv \frac{c_1}N$, so that the probability of at
least one bad avalanche between $\theta+1$ and $t$ is bounded by $\frac{K c_1}N$.
We conclude that for $t \in \{\theta+1, \ldots, \theta+K\}$,
$\prob^N_\mathbf{0}(A_{\theta j}(t)>\epsilon) \leq \frac{Kc_1 + c_2}N \leq \frac {cK}N$,
for some $c>0$.
\qed

\medskip\noindent
{\it Proof of Lemma \ref{eentheta}, part (2).}
From Lemma \ref{alpha}, it follows that if $G_t(\alpha)$ occurs, and after $s$
time steps all $\theta$-heavy sites have toppled at least once, in avalanches that all
start at least a distance $\lceil \frac 1\alpha \rceil$ from all current $\theta$-heavy
sites, then $G_{t+s}(f(\alpha))$ occurs. We will exploit this fact as follows.

Suppose that $G_t(\alpha)$ occurs and that in addition, at time $t$ the distribution of the empty set - if present - is uniform on $\{1,\ldots,N\}$. We claim that this implies that
$G_{t+2}(f(\alpha))$ occurs with a probability that is bounded below, uniformly in $N$ and
$\epsilon$. To see this, observe that
if there is no empty site, then all $N$ sites topple in one time step. If there is an empty site, this
(meaning all sites topple)
also happens in two steps if in the first step, we add to one side of the empty site, and in the second
step to the other side. Denote by $e_1$ the position of the empty site before the first
addition (at $X_{t+1}$), and $e_2$ before the second addition (at $X_{t+2}$).
If $X_{t+1} < e_1$, then $e_2<e_1$. Therefore, all sites topple if $X_{t+1} <
e_1$ and $X_{t+2} > e_1$. With the distribution of $e_1$ uniform on
$\{1,\ldots,N\}$, the probability that this happens is bounded below by some constant
$\gamma'$ independent of $N$ and $\epsilon$. However we have the extra demand that both additions should start
at least a distance $\lceil \frac 1\alpha \rceil \leq \lceil \frac 1\epsilon \rceil$
from all current $\theta$-heavy sites, of which there are at most $\lceil \frac
1\epsilon \rceil$. Thus, both additions should avoid at most $\lceil \frac 1\epsilon
\rceil^2$ sites. The probability that this happens is therefore less than some $\gamma'>0$, but it is
easy to see that the difference decreases with $N$. We can then conclude that there is an
$N'$ large enough so that the probability that this happens is at least $\gamma > 0$ for all $N
\geq N'$, with $0<\gamma<1$ independent of $N$ and $\epsilon$.

In view of this, the probability that $G_{\theta+2}(f(1))$ occurs, for $N$ large enough, is at
least $\gamma$. We wish to iterate this argument $w$ times. However, the lower bound
$\gamma$ is only valid when the distribution of the empty site, if present, is uniform on
$\{1,\ldots,N\}$. We do not have this for $\eta(\theta+2)$ since we have information about what
happened in the time interval $(\theta,\theta+2)$. However, after one more addition, the position
of the empty site in $\eta(\theta+3)$, if present, is again uniform on
$\{1,\ldots,N\}$. Since $f^w(1) \leq \epsilon$, iterating this argument gives
$$
\prob^N_\mathbf{0}(\max_j A_{\theta j}(t) > \epsilon) \leq (1-\gamma)^{t-\theta-3w}.
$$
\qed

\medskip\noindent
{\it Proof of Theorem \ref{quasiunits}, part (2).}
By Lemma \ref{variantieherschrijven}, it suffices to prove \eqref{alleenmaardit}.
We estimate, using that $\sum_{\theta}A_{\theta j_N}(t) \leq 1$,
$$
\expec^N_\mathbf{0}\left[\sum_{\theta=1}^t (A_{\theta j_N}(t))^2\right]  \leq
\expec^N_\mathbf{0}\left[\max_{1\leq\theta\leq t} A_{\theta j_N}(t) \sum_{\theta=1}^t
A_{\theta j_N}(t)\right] \leq  \expec^N_\mathbf{0}\left[\max_{1\leq\theta\leq t}
A_{\theta j_N}(t)\right]
$$
$$
 \leq  \expec^N_\mathbf{0}\left[\max_{t-K \leq \theta \leq t}A_{\theta j_N}(t)\right] +
\expec^N_\mathbf{0}\left[\max_{N(N+1) < \theta < t-K}A_{\theta j_N}(t)\right] +
\expec^N_\mathbf{0}\left[\max_{1 \leq \theta \leq N(N+1)}A_{\theta j_N}(t)\right].
 $$
We then for the first two terms estimate, using that $\max_\theta A_{\theta j_N}(t)
\leq 1$,
$$
\expec^N_\mathbf{0}[\max_\theta A_{\theta j_N}(t)] \leq \epsilon +
\prob^N_\mathbf{0}(\max_\theta A_{\theta j_N}(t) > \epsilon) \leq \epsilon +
\sum_\theta \prob^N_\mathbf{0}(A_{\theta j_N}(t)>\epsilon).
$$
We finally use Lemma \ref{eentheta}, and choose $K = K_N$ increasing with $N$. For
$\theta \in [t,t-K_N]$, we straightforwardly obtain $\sum_{\theta=t-K_N}^t
\prob^N_\mathbf{0}\left(A_{\theta j_N}(t)>\epsilon\right)  = O(\frac {K_N^2}N)$,
uniformly in $t$, as $N \to \infty$.
For $\theta < t-K_N$ we calculate
$$
\sum_{\theta < t-K_N} \prob^N_\mathbf{0}\left(A_{\theta j_N}(t)>\epsilon\right) \leq
\sum_{t-\theta>K_N}(1-\gamma)^{t-\theta-3w} = O((1-\gamma)^{K_N}), \hspace{2.8cm} N \to
\infty,
$$
so that
$$
\expec^N_\mathbf{0}\left[\sum_{\theta=1}^t(A_{\theta j_N}(t))^2\right] \leq 2\epsilon +
O(\frac {K_N^2}N) + O((1-\gamma)^{K_N}) + \expec^N_\mathbf{0}\left[\max_{1 \leq \theta
\leq N(N+1)}A_{\theta j_N}(t)\right], \hspace{1cm} N\to\infty.
$$
In the limit $t \to \infty$, by Lemma \ref{uitsmeren} part (2), the last term vanishes.
We now choose $K_N = N^{1/3}$, to obtain
%\sqrt[3]{N}
$$
\limsup_{N\to\infty}\lim_{t\to\infty}\expec^N_\mathbf{0}\left[\sum_{\theta=1}^t (A_{\theta
j_N}(t))^2\right] \leq 2\epsilon,
$$
Since $\epsilon>0$ is arbitrary, we finally conclude that
$$
\lim_{N\to\infty}\lim_{t\to\infty}\expec^N_\mathbf{0} \left[\sum_{\theta=1}^t
(A_{\theta j_N}(t))^2\right] = 0.
$$
\qed

\section{The $(N,[0,1])$-model}
\label{nultoteensection}

\subsection{Uniqueness of the stationary distribution}

\begin{theorem}
The $(N,[0,1])$ model has a unique stationary distribution $\upsilon_N$. For every
initial distribution $\nu$ on $\Omega_N$, $\prob_\nu$ converges in total variation to
$\upsilon_N$.
\label{nu}
\end{theorem}

\begin{proof}
We prove this theorem again by constructing a successful coupling. For clarity, we
first treat the case $N=2$, and then generalize to $N > 2$. The coupling is best
described in words.

Using the same notation as in previous couplings, we call two independent copies of the
process $\eta^1(t)$ and $\eta^2(t)$, and call the coupled processes $\hat{\eta}^1(t)$
and $\hat{\eta}^2(t)$. Initially, we choose $\hat{\eta}^1(t) = \eta^1(t)$, and
$\hat{\eta}^2(t) = \eta^2(t)$. It is easy, but tedious, to show that
$\eta^1_1(t)=\eta^2_1(t)=0$, while $\re(\eta^1_2(t))=\re(\eta^2_2(t)) = 1$, occurs
infinitely often.

At the first such time $T_1$ that this occurs, we choose the next addition as follows.
Call $\Delta(t) = \eta^1_2(t) - \eta^2_2(t)$. We choose $\hat{X}^2_{T_1+1} =
X^1_{T_1+1}$, and $\hat{U}^2_{T_1+1} = (U^1_{T_1+1}+\Delta(T_1))\mbox{mod}~1$. Observe
that the distribution of $\hat{U}^2_{T_1+1}$ is uniform on $[0,1]$.

This addition is such that with positive probability the full sites are chosen for the
addition, and the difference $\Delta(T_1)$ is canceled. More precisely, this occurs if
$X^1_{T_1+1} = 2$, which has probability 1/2, and
$(U^1_{T_1+1}+\Delta(T_1))\mbox{mod}~1 = U^1_{T_1+1}+\Delta(T_1)$, which has
probability at least 1/2, since $\eta^1_2(T_1)$ and $\eta^2_2(T_1)$ are both full,
therefore $\Delta(T_1) \leq 1/2$. If this occurs, then we achieve success, i.e.,
$\hat{\eta}^1(T_1+1)=\hat{\eta}^2(T_1+1)$, and from that time on we can let the two
coupled processes evolve together.

If $\hat{\eta}^1(T_1+1) \neq \hat{\eta}^2(T_1+1)$, then we evolve the two coupled
processes independently, and repeat the above procedure at the next instant that
$\hat{\eta}^1_1(t)=\hat{\eta}^2_1(t)=0$. Since at every such instant, the probability
of success is positive, we only need a finite number of attempts. Therefore, the above
constructed coupling is successful, and this proves the claim for $N=2$.

We now describe the coupling in the case $N>2$. We will again evolve two processes
independently, until a time where $\eta^1_1(t) = \eta^2_1(t) = 0$, while all other
sites are full. At this time we will attempt to cancel the differences on the other
$N-1$ sites one by one. We define $\Delta_j(t) = \eta^1_j(t)-\eta^2_j(t)$, and as
before we would be successful if we could cancel all these differences. However, now
that $N>2$, we do not want an avalanche to occur during this equalizing procedure,
because we need $\eta^1_1(t) = \eta^2_1(t) = 0$ during the entire procedure.
Therefore, we specify $T_1$ further: $T_1$ is the first time where not only
$\eta^1_1(t) = \eta^2_1(t) = 0$ and all other sites are full, but also $\eta^1_j(t) <
1-\epsilon$ and $\eta^2_j(t) < 1- \epsilon$, for all $j = 2, \ldots, N$, with
$\epsilon=\frac 1{2^{N+1}}$. At such a time, a positive amount can be added to each
site without starting an avalanche. We will first show that this occurs infinitely
often, which also settles the case $N=2$.

By Proposition \ref{compare}, after a finite time $\eta^1(t)$ and $\eta^2(t)$ contain
at most one non-full site. It now suffices to show that for any $\xi(t) \in \Omega_N$
with at most one non-full site, with positive probability the event that
$\xi_1(t+4)=0$, while $\xi_j(t+4)\leq 1-\frac 1{2^{N+1}}$ for every $2 \leq j \leq N$,
occurs.

One explicit possibility is as follows. The first addition should cause an avalanche.
This will ensure that $\xi(t+1)$ contains one empty site. This occurs if the addition
site is a full site, and the addition is at least $1/2$. The probability of this is at
least $\frac 12 (1-\frac 1N)$.
The second addition should change the empty site into full. For this to occur, the
addition should be at least 1/2, and the empty site should be chosen. This has
probability $\frac 1{2N}$.
The third addition should be at least 1/2 to site $1$, so that an avalanche is started
that will result in $\xi_N(t+3)=0$. This has again probability $\frac 1{2N}$.
Finally, the last addition should be an amount in $[\frac 12, \frac 34]$, to site
$N-1$. Then by (\ref{avalanche}), every site but site $N$ will topple once, and after
this avalanche, site 1 will be empty, while every other site contains at most $1-\frac
1{2^{N+1}}$. This last addition has probability $\frac 1{4N}$.

Now we show that at time $T_1$ defined as above, there is a positive probability of
success.
To choose all full sites one by one, we require, first, for all $j = 2, \ldots, N$ that
$X^1_{T_1+j-1} = j$. This has probability $(\frac 1N)^{N-1}$. Second, we need
$(U^1_{T_1+j-1}+\Delta_j(T_1))\mbox{mod}~1 = U^1_{T_1+j-1}+\Delta_j(T_1)$ for all $j =
2, \ldots, N$. This event is independent of the previous event and has probability at least $(\frac 12)^{N-1}$. If this second
condition is met, then third, we need to avoid avalanches, so for all $j = 2, \ldots,
N$, $\eta^1_j(T_1+j-1)+U^1_{T_1+j-1} = \eta^2_j(T_1+j-1)+\hat{U}^2_{T_1+j-1} <1$. It is not
hard to see that this has positive conditional probability, given the previous events.
We conclude that the probability of success at time $T_1+N-1$ is positive, so that we
only need a finite number of such attempts. Therefore, the coupling is successful, and
we are done.
\end{proof}

\subsection{Simulations}

\begin{figure}[ht]
 \centerline{\includegraphics[width=7cm]{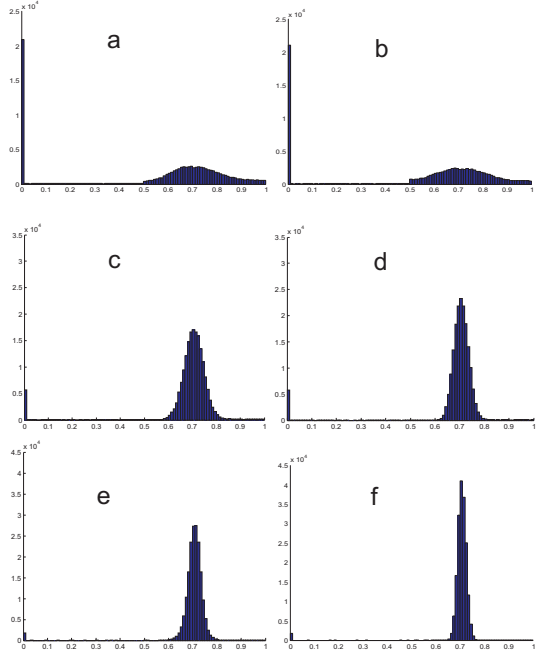}}
 \caption{Simulation results for the $(N,[0,1])$-model. The histograms represent
observed energies during 100,000 (a,b) and 200,000 iterations (c-f). The system size
was 3 sites (a,b), 30 sites (c,d) and 100 sites (e,f). (a),(c) and (e) are boundary
sites, (b), (d) and (f) are central sites.}
 \label{zhangsims}
\end{figure}

We performed Monte Carlo simulations of the $(N,[0,1])$-model, for various values of
$N$. Figure \ref{zhangsims} shows histograms of the energies that a site assumes during all
the iterations. We started from the empty configuration, but omitted the first 10$\%$
of the observations to avoid recording transient behavior. Further increasing this
percentage, or the number of iterations, had no visible influence on the results.

The presented results show that, as the number of sites of the model increases, the
energy becomes more and more concentrated around a value close to 0.7. In the next section,
we present an argument for this value to be $\sqrt{1/2}$.
We further observe that it seems to make a difference where the site is located: at the
boundary the variance seems to be larger than in the middle.

\subsection{The expected stationary energy per site as $N \to \infty$}
\label{expectedsection}

From the simulations it appears that for large values of $N$,
the energy per site concentrates at a value close to 0.7, for every site. Below we argue,
under some assumptions that are consistent with our simulations, that this value
should be $\sqrt{1/2}$.

First, we assume that every site has the same expected stationary energy.
Moreover, we assume that pairs of sites are asymptotically independent, i.e.,
$\eta_x$ becomes independent of $\eta_y$ as $|x-y| \to \infty$. (If the stationary
measure is indeed such that the energy of every site is a.s.\ equal to a constant, then
this second assumption is clearly true.)
With $\expec_{\upsilon_N}$ denoting expectation with respect to the stationary
distribution $\upsilon_N$, we say that $(\upsilon_N)_N$ is {\em asymptotically independent} if for any
$1 \leq x_N, y_N \leq N$ with $|x_N-y_N| \to \infty$, and for any
$A,B$ subsets of $\mathbb{R}$ with positive Lebesgue measure, we have
\beq
\lim_{N \to \infty} \left(
\expec_{\upsilon_N}({\bf 1}_{\eta_{x_N}\in B}|\eta_{y_N}\in A)-
\expec_{\upsilon_N}({\bf 1}_{\eta_{x_N}\in A}\right) = 0.
\label{expected2}
\eeq

\begin{theorem}
Suppose that in the $(N,[0,1])$ model, for any sequence $j_N \in \{1,\ldots,N\}$,
\beq
\lim_{N \to \infty} \expec_{\upsilon_N} (\eta_{j_N}) = \rho,
\label{expected1}
\eeq
for some constant $\rho$. Suppose in addition that $(\upsilon_N)_N$ is asymptotically independent.
Then we have $\rho = \sqrt{\frac 12}$.
\end{theorem}

\begin{proof}
The proof is based on a conservation argument. If we pick a configuration according to
$\upsilon_N$ and we make an addition $U$, we denote the random amount that leaves the system
by $E_{out,N}$. By stationarity, the expectation of $U$
must be the same as the expectation of $E_{out,N}$.

The amount of energy that leaves the system in case of an avalanche, depends on
whether or not one of the sites is empty (or behaves as empty). Remember (Proposition
\ref{compare}) that when we pick a configuration according to the stationary distribution, then
there can be at most one empty or
anomalous site. If there is one empty site, then the avalanche reaches one boundary. If
there are only full sites, then the avalanche reaches both boundaries, and in case of
one anomalous site, both can happen.

However, configurations with no empty site have
vanishing probability as $N \to \infty$: we claim that the stationary
probability for a configuration to have no empty site, is bounded above by $p_N$, with
$\lim_{N\to\infty}p_N = 0$.
To see this, we divide the support of the stationary distribution into two sets:
$\mathcal{E}$, the set of configurations with one empty site, and $\mathcal{N}$, the
set of configurations with no empty site. The only way to reach $\mathcal{N}$ from
$\mathcal{E}$, is to make an addition precisely at the empty site. As $X$ is uniformly
distributed on $\{1,\ldots,N\}$, this has probability $\frac 1N$, irrespective of the
details of the configuration. The only way to reach
$\mathcal{E}$ from $\mathcal{N}$, is to cause an avalanche; this certainly happens if
an addition of at least $1/2$ is made to a full site. Again, since $X$ is uniformly
distributed on $\{1,\ldots,N\}$, and since there is at most one non-full site, this has
probability at least $\frac 12 \frac {N-1}{N}$.

Now let $X$ be the (random) addition site at a given time, and denote by $A_x$ the event that
$X=x$ and that this addition causes the start of an avalanche.
Since $E_{out,N}=0$ when no avalanche is started, we can write
\beq
\expec_{\upsilon_N}(E_{out,N}) = \sum_{x=1}^N
\expec_{\upsilon_N}(E_{out,N}|A_x)\prob_{\upsilon_N}(A_x).
\label{uit}
\eeq
We calculate $\prob_{\upsilon_N}(A_x)$ as
follows, writing $U$ for the value of the addition:
\begin{eqnarray}
\label{ggg}
\prob_{\upsilon_N}(A_x) & = & \frac{1}{N}\prob_{\upsilon_N}(\eta_x+U \geq 1)
= \frac{1}{N}\prob_{\upsilon_N}(U \geq 1-\eta_x) \nonumber \\
& = & \frac{1}{N}\int\prob_{\upsilon_N}(U \geq 1-\eta_x)d\upsilon_N(\eta)
=\frac{1}{N} \int \eta_x d\upsilon_N(\eta) \nonumber \\
& = & \frac{1}{N}\expec_{\upsilon_N}(\eta_x).
\label{probA}
\end{eqnarray}
Let $L_N = \lceil \log N \rceil$. Even if the
avalanche reaches both boundary sites, the amount of energy that leaves the system can
never exceed 2, which implies that
\begin{equation}
\label{uu}
\left|\sum_{x=1}^N \expec_{\upsilon_N}(E_{out,N}|A_x) -
\sum_{x=2L_N}^{N-2L_N} \expec_{\upsilon_N}(E_{out,N}|A_x)\right| \leq 8L_N.
\end{equation}
It follows from (\ref{uit}), (\ref{ggg}) and (\ref{uu}) that
\begin{equation}
\label{voorlopig}
\expec_{\upsilon_N}(E_{out,N}) = \frac{1}{N}\sum_{x=2L_N}^{N-2L_N}
\expec_{\upsilon_N}(E_{out,N}|A_x)\expec_{\upsilon_N}(\eta_x) + O(L_N/N).
\end{equation}

If the avalanche, started at site $x$, reaches the boundary at site 1, then the amount
of energy that leaves the system is given by $\frac 12 \eta_1 + \frac 14 \eta_2 +
\cdots + (\frac 12)^x (\eta_x +U)$.
For all $x \in \{2L_N,\ldots,N-2L_N\}$, this can be written as
$$
\frac 12 \eta_1 + \frac 14 \eta_2 + \cdots + (\frac 12)^{L_N}\eta_{L_N} + (\frac
12)^{L_N+1}\eta_{L_N+1} + \cdots + (\frac 12)^x (\eta_x +U),
$$
where for the last part of this expression, we have the bound
$$
(\frac 12)^{L_N+1}\eta_{L_N+1} + \cdots + (\frac 12)^x (\eta_x +U) \leq (\frac
12)^{L_N}.
$$
Since the occurrence of $A_x$ depends only on $\eta_x$ (and on $X$ and $U$), for $2L_N \leq x \leq
N-2L_N$, by asymptotic independence
there is an $\alpha_N$, with $\lim_{N\to\infty} \alpha_N = 0$, such that for all $1
\leq i \leq L_N$ and $ 2L_N \leq x \leq N-2L_N$, we have
$$
|\expec_{\upsilon_N}(\eta_i|A_x) - \expec_{\upsilon_N}(\eta_i)| \leq \alpha_N,
$$
so that
$$
\left|\expec_{\upsilon_N}\left(\frac 12 \eta_1 + \cdots + (\frac
12)^{L_N}\eta_{L_N}|A_x\right)-\expec_{\upsilon_N}\left(\frac 12 \eta_1 + \cdots +
(\frac 12)^{L_N}\eta_{L_N}\right)\right| \leq \left(\frac 12 + \frac 14 + \cdots\right)\alpha_N,
$$
which is bounded above by $\alpha_N$. By symmetry, we have a similar result in the case that the other boundary is reached.
In case both boundaries are reached, we simply use that the amount of energy
that leaves the system is bounded above by 2.

In view of this, we continue the bound in (\ref{voorlopig}) as follows
\begin{eqnarray*}
\nonumber \expec_{\upsilon_N}(E_{out,N}) & = & \frac 1N \sum_{x=2L_N}^{N-2L_N}\left(
\expec_{\upsilon_N}(\frac 12 \eta_1 + \frac 14 \eta_2 + \cdots + (\frac
12 )^{L_N}\eta_{L_N}| A_x) +O((\frac{1}{2})^{L_N}) \right)\expec_{\upsilon_N}(\eta_x ) + \\
& & + O(L_N/N)\\
& = & \frac 1N \sum_{x=2L_N}^{N-2L_N}
\expec_{\upsilon_N}(\frac 12 \eta_1 + \frac 14 \eta_2 + \cdots + (\frac
12 )^{L_N}\eta_{L_N})\expec_{\upsilon_N}(\eta_x ) + \\
& & + O(\frac{L_N}{N}) + O((\frac12 )^{L_N}) + O( \alpha_N ) + O(p_N),
\end{eqnarray*}
as $N \to \infty$.
Letting $N \to \infty$ and inserting \eqref{expected1} now gives
$$
\lim_{N\to\infty}\expec_{\upsilon_N}(E_{out,N}) = \rho^2.
$$
As the expectation of $U$ is $\frac 12$, we conclude that $\rho = \sqrt{\frac 12}$.
\end{proof}


\begin{thebibliography}{}
\bibitem{cessac}
{P. Blanchard, B. Cessac, T. Kr\"{u}ger}:
{A Dynamical System Approach to SOC Models of Zhang's Type}.
{\em Journ. of Stat. Phys.} {\bf 88}(1/2), 307-318 (1997)

\bibitem{dhar}
{D. Dhar}:
{Self-Organized Critical State of Sandpile Automaton Models}.
{\em Phys. Rev. Lett.} {\bf 64}(14), 1613-1616 (1990)

\bibitem{delay}
{O. Diekmann, S.A. van Gils, S.M. Verduyn Lunel and H.-O. Walther}:
{Delay equations; Applied Mathematical Sciences vol. 110}.
{\em Springer-Verlag,} New York, 1995

\bibitem{dudley}
{R.M. Dudley}:
{Real Analysis and Probability}.
{\em Wadsworth \& Brooks/Cole,} California, 1989

\bibitem{feller}
{W. Feller}:
{An Introduction to Probability Theory and its Applications, vol. II}.
{\em John Wiley and Sons, Inc.,} USA, 1966

\bibitem{janosi}
{I.M. Janosi}:
{Effect of anisotropy on the self-organized critical state}.
{\em Phys. Rev. A} {\bf 42}(2), 769-774 (1989)

\bibitem{meester}
{R. Meester, F. Redig and D. Znamenski}:
{The abelian sandpile model; a mathematical introduction}.
{\em Markov Proc. Rel. Fields} {\bf 7}, 509-523 (2001)

\bibitem{pastor}
{R. Pastor-Satorras and A. Vespignani}:
{Anomalous scaling in het Zhang model}.
{\em Eur. Phys. J. B } {\bf 18}, 197-200 (2000)

\bibitem {priezzhev}
{E.V. Ivashkevich and V.B. Priezzhev}:
Introduction to the sandpile model.
{\em Physica A} {\bf 254}, 97--116 (1998)

\bibitem{redig}
{C. Maes, F. Redig, E. Saada and A. van Moffaert}:
{On the thermodynamic limit for a one-dimensional sandpile process}.
{\em Markov Proc. and Rel. Fields} {\bf 6}(1), 1-21 (2000)

\bibitem{thorisson}
{H. Thorisson}:
{Coupling, Stationarity, and Regeneration}.
{\em Springer Verlag,} New York, 2000

\bibitem{turcotte}
{D.L. Turcotte}:
{Self-organized criticality}.
{\em Reports on progress in physics} {\bf 62} (10), 1377-1429 (1999)

\bibitem{zhang}
{Y.-C. Zhang}:
{Scaling theory of Self-Organized Criticality}.
{\em Phys. Rev. Lett.} {\bf 63}(5), 470-473 (1989)
\end{thebibliography}
\end{document}